\documentclass[english,aps,prstper,reprint,showpacs,titlepage,longbibliography]{revtex4-2}

\usepackage[T1]{fontenc}	
\usepackage[latin9]{inputenc}	
\usepackage{geometry}		
\usepackage{graphicx}
\usepackage[above,below]{placeins}	
\usepackage{times}
\usepackage{amsmath}
\usepackage{physics}
\usepackage{array}
\usepackage{booktabs}
\usepackage{multirow}

\usepackage[normalem]{ulem}
\useunder{\uline}{\ul}{}
\usepackage{longtable}

\usepackage{tagging}
\newcommand{\todo}[1]{\tagged{todo}{\textbf{TODO #1}}}
\newcommand{\rem}[1]{\tagged{removed}{\st{#1}}}

\newcommand{\hl}[1]{#1}

\usepackage{enumitem}          
\setlist{nosep}                 
\usepackage{hyperref}
\hypersetup{colorlinks=true,urlcolor=blue,citecolor=blue,linkcolor=blue}   
\newcommand{\drc}{Cs\'{\i}kszentmih\'{a}lyi\ }

\newcommand{\etal}{\textit{et al.}\ }

\newcommand{\rqone}{What changes are observed in students' epistemology about experimental physics as a result of the NOMR labs?\ }
\newcommand{\rqtwo}{What changes are observed in students' physics self-efficacy in experimental physics as a result of the NOMR labs?\ }

\newcommand{\rqthree}{To what extent are students productively engaged in the NOMR activities, and how does that engagement compare with the hands-on labs in the same course?\ }

\begin{document}

\title{Modeling novel physics in virtual reality labs: An affective analysis of student learning}%

\author{Jared P. Canright}%
\email{jpcan@uw.edu}
\author{Suzanne White Brahmia}
\email{brahmia@uw.edu}
\affiliation{Department of Physics, University of Washington, 3910 15th Ave NE, Seattle, WA, 98195}
\date{\today}%
\begin{abstract}
    We report on a study of the effects of laboratory activities that model fictitious laws of physics in a virtual reality environment on (1) students' epistemology about the role of experimental physics in class and in the world; (2) students' self-efficacy; and (3) the quality of student engagement with the lab activities. 
    We create opportunities for students to practice physics as a means of creating and validating new knowledge by simulating real and fictitious physics in virtual reality (VR). 
    This approach seeks to steer students away from a confirmation mindset in labs by eliminating any form of prior or outside models to confirm. We refer to the activities using this approach as Novel Observations in Mixed Reality (NOMR) labs. 
    We examined NOMR's effects in 100-level and 200-level undergraduate courses.
    Using pre-post measurements we find that after NOMR labs, students in both populations were more expertlike in their epistemology about experimental physics and held stronger self-efficacy about their abilities to do the kinds of things experimental physicists do. 
    Through the lens of the psychological theory of flow, we found that students engage as productively with NOMR labs as with traditional hands-on labs. This engagement persisted after the novelty of VR in the classroom wore off, suggesting that these effects are due to the pedagogical design rather than the medium of the intervention.
    We conclude that these NOMR labs offer an approach to physics laboratory instruction that centers the development of students' understanding of and comfort with the authentic practice of science.
\end{abstract}
\maketitle

\section{Introduction}

This study seeks to characterize aspects of student learning that are both highly valued \cite{AAPTLabGuidelines} and challenging to assess. In the context of experimental physics courses and using a virtual reality (VR) environment, students engage in activities with novel force laws that are designed to meet a need for introductory laboratory activities that deepen undergraduate physics students' understanding of the process of generating \textit{new }knowledge in science, and the quantitative scientific scrutiny involved. The objective of this study is to better understand the extent to which exploring novel physics, made possible through the use of immersive technologies, can render students more expert-like in their beliefs (1) about how scientific knowledge is generated and (2) in their capacity to produce scientific knowledge. 

In light of the physics education research community's current understanding that laboratory instruction is not an effective means of teaching conceptual content \cite{Smith2021BestLabs, Holmes2017ValueContent}, we instead seek to use labs as a place where students engage in the authentic practice of science, equipping them with the tools to understand the world through an empirical lens in alignment with the AAPT's recommendations for undergraduate physics lab instruction \cite{AAPTLabGuidelines}. We aim to foster an expertlike understanding of the role and process of experimentation \cite{HuPRPER2017, Etkina2007InvestigativePhysics}, build their confidence in their ability to design, perform, and interpret experiments \cite{Bandura1977Self-efficacy:Change., Etkina2006ScientificAssessment}, and keep them actively engaged through the whole process \cite{Freeman2014ActiveMathematics}.

Understanding how to provide engaging opportunities for students to develop mathematical models of novel phenomena in a teaching laboratory is a difficult open problem in physics pedagogy \cite{AAPTLabGuidelines, Smith2021BestLabs, Trumper2003ThePerspectives}.
This kind of divergent, creative activity is fraught with challenges related to student autonomy and safety, \todo{P4 I think there's a relevant Holmes paper here?} opportunities for meaningful contexts \cite{Trumper2003ThePerspectives}, and the expertise of instructors to engage in a manner that responds to what is happening within each group. \todo{P4 cite Doucette?} These issues are particularly challenging in large-enrollment courses where labs are commonly taught by inexperienced undergraduate and graduate teaching assistants (TAs).

Our framework for designing activities in which students learn to generate models is the Investigative Science Learning Environment (ISLE) approach  \cite{Etkina2007InvestigativePhysics, Brookes2020ImplementingLearning, Etkina2021ThePhysics, Etkina2015MillikanPractices, Etkina2015DefiningPhenomenon, isle2020} by Etkina, Brookes, and Planinsic. In ISLE, model generation happens during what the authors have named  \textit{observational experiments}, where students engage in open-minded exploration with the goal of developing a model for an unknown phenomenon. This phase of the ISLE process, itself a simplified but authentic representation of the scientific process, is followed by iteratively testing, revising, refining, and applying the model. In our approach, we alter the language slightly from Etkina to optimize transparency for the students of what they are doing. We refer to the processes of \textit{model-generating} and \textit{model-testing}, rather than observational and testing, experiments.     Note that, in context of ISLE, the terms ``model,'' ``explanation,'' and ``hypothesis'' are interchangeable \cite{Etkina2006TheInstruction}.

Model-generating experiments involving novel scenarios are difficult to create, especially in introductory courses. In many cases, experimental physics questions that reasonably could be investigated at the introductory level are well-known with easily-Googled answers. In presence of known answers, students tend to hold those in highest regard, seeking to confirm known answers above anything and everything else, even in the face of contradictory data \cite{Stein2018ConfirmingLabs} or explicit instruction to the contrary. Thus, any access to a ``right answer'' can derail efforts to engage students in authentic model generation. We address this expectation of getting a right answer by putting students into a different universe with \textit{new }physics that builds on Newton's laws and fundamental conservation laws -- where neither they, nor their textbook, nor Google, nor their TAs have a ready-made model at hand. The problem of shifting students' mindset toward generating new models becomes trivial when there are no existing models to confirm.

The activities described in this paper have been part of the University of Washington (UW) introductory physics curriculum for over two years. In the   Novel Observations in Mixed Reality (NOMR) labs \cite{Canright2020,Canright2021DevelopingReality}, students explore real and fictitious physical phenomena in an immersive 3D environment. Instructors are struck by the ways that students mature as scientists through these labs, an impression that is not easily quantified. 
The following is an excerpt from a post-course survey that is fairly typical at the sophomore level, and representative of those most impacted at the introductory level.  

\begin{quote}
    ``VR labs were fantastic for learning how to effectively approach a physics situation where I didn't already know what would happen. In most
experiments I have done in previous courses, I had learned what to expect before I was actually making observations and collecting data, so this
course helped me learn a new way to approach experiments.''
\end{quote}


 This study is a step toward characterizing this kind of intellectual growth we observe in many students. The work is situated in efforts across the physics education community to find and adapt affective assessment tools beyond standard course evaluations. Our study seeks to establish whether students' belief in their ability to do physics and their sense of belonging in physics grow along with their understanding of the nature and role of experimental physics to generate new knowledge. We assess the impact of the intervention on students': epistemology about experimental physics; physics \rem{identity and }self-efficacy; and engagement in the learning process.  This study contributes to the ongoing research into assessment of what students take away from effective laboratory instruction \cite{Walsh2019QuantifyingThinking, Vignal2023SurveyLabs}.
Specifically, we focus on 
the following research questions:
\begin{enumerate}[label=\textbf{RQ\arabic*}]
    \item\label{rq:1} \rqone
    \item\label{rq:2} \rqtwo 
    \item\label{rq:3} \rqthree
\end{enumerate}

\section{Background}


\subsection{ISLE}

We begin by considering what lab activities reflecting the real-world practice of science look like. The ISLE approach to physics education \cite{Etkina2007InvestigativePhysics, Brookes2020ImplementingLearning, Etkina2021ThePhysics, Etkina2015MillikanPractices, Etkina2015DefiningPhenomenon, isle2020} prioritizes epistemologically authentic investigation of physics as a means to develop students' scientific abilities \cite{Etkina2006ScientificAssessment} and habits of mind. Teaching students to think like expert physicists takes priority over covering conceptual content.

\begin{figure}[htbp]
    \centering
    \includegraphics[width=0.5\textwidth]{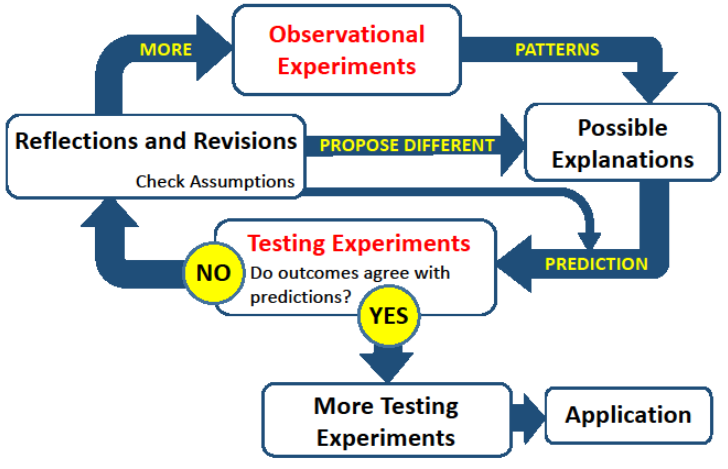}
    \caption{The ISLE process is a simplified representation of the real-world practice of science, iteratively generating, testing, rejecting, and refining models to empirically create and validate knowledge.}
    \label{fig:isle}
\end{figure}

Three types of experiments form the core of ISLE instructional activities, related to each other by the ISLE process (Figure \ref{fig:isle}):
\begin{description}
    \item[Model-generating experiments] Labelled in the diagram as Observational Experiments, students engage in open-minded exploration of a previously unknown physical phenomenon. They make note of patterns in the phenomenon's behavior and devise explanations for those patterns. These patterns become mathematical models and mechanistic explanations of the phenomenon. The models created in model-generating experiments form the basis of students' knowledge of the phenomenon.
    \item[Testing experiments] A model from a prior model-generating experiment is tested. Students design an experiment with well-determined independent, dependent, and controlled variables to test a prediction about the outcome of an experiment that follows from the model in question. They run the experiment, collect and analyze their data, and judge whether the outcome is consistent with the prediction. If so, they have supported, or failed to reject, the model. If not, the model is rejected.
    \item[Application experiments] Students apply tested models to determine the value of unknown physical quantities or solve practical problems.
\end{description}

In the ISLE approach, content and process are considered to be inextricably paired; these three types of activities are how students uncover new physics content. Students encounter physical phenomena for the first time through hands-on experimentation, and only after identifying patterns and developing their own explanations do they read about the phenomena in their text.

ISLE-approach labs consist primarily of questions to guide students' thoughts rather than dictate them. In this way, students learn to ask and answer questions in the way that a scientist would. This process is guided and refined by scientific abilities rubrics \cite{Etkina2006ScientificAssessment} used to assess and give feedback on their work.

\subsection{Affective measures} 

This study employs three different research-validated surveys to explore different aspects of students' engagement and learning. Table \ref{tab:instruments} gives the name, abbreviation, target metrics, format, and administration schedule for each survey.

\begingroup
\squeezetable
\begin{table}[htbp]
\begin{tabular}{|p{0.14\linewidth}|p{0.05\linewidth}|p{0.3\linewidth}|p{0.225\linewidth}|p{0.225\linewidth}|}
\multicolumn{2}{c}{\textbf{Survey}}                & \multicolumn{1}{c}{\textbf{Target Metrics}} & \multicolumn{1}{c}{\textbf{Format}}   & \multicolumn{1}{c}{\textbf{Schedule}} \\ \hline
Lab Epistemology Survey 
    & LES 
    & Student attitudes and beliefs about the nature and role of experimentation in physics    
    & Five open-ended short-answer questions 
    & \multirow{2}{\linewidth}{Pre-survey at the beginning of the term;\par 100-level post-survey after NOMR activities concluded; 200-level post-survey after course final}  \\ \cline{1-4} 

Physics Identity Survey & PhIS & Degree of students' self-identification as a physicist, interest in physics, and belief in their ability to practice and succeed at physics   & Six 6-point Likert items (self-perception and interest); \par Five 7-point Likert items (self-efficacy) &                                                                                                                                                                                                                     \\ \hline
Flow Survey             & FS & Degree and nature of students' engagement with the week's lab activity     & Seven 7-point Likert items                                                                                      & Weekly at the conclusion of each lab activity                                                                                                                                                                                                                                            \\ \bottomrule
\end{tabular}
\caption{The characteristics and administration schedule of each survey used in this study are summarized.}
\label{tab:instruments}
\end{table}
\endgroup

\subsubsection{\label{sec:les_background}Epistemology about experimental physics}

The Lab Epistemology Survey (LES) originally developed by Hu and Zwickl \cite{HuZwickl2017, Hu2018ExaminingStudents} is used in this study as a measure of students' epistemology about the role of experimental physics in class and in the world. The LES was developed as an instrument to characterize the beliefs held by physics students at introductory undergraduate, upper-division undergraduate, and graduate levels about about experimentation, models, and their roles in the scientific process.  This study uses the LES pre-post to assess changes in students' epistemology about experimental physics before and after students complete the NOMR labs. Measuring changes in students' epistemology gives us a window into whether they are adopting the beliefs, attitudes, and mindset about experimental physics characteristic of expert physicists.

The LES is composed of six open-ended questions, accompanied by a codebook used to identify themes in student responses in a consistent and reproducible way. We focus on the first two questions:
\begin{enumerate}[label=\textbf{LES\arabic*}]
    \item\label{q:1} In your opinion, why are experiments a common part of physics classes? Provide examples or any evidence to support your answer.
    \item\label{q:2} In your opinion, why do scientists do experiments for their research? Provide examples or any evidence to support your answer.
\end{enumerate}

\newcolumntype{P}[1]{>{\centering\arraybackslash}p{#1}}
\newcommand{\centered}[1]{\begin{tabular}{l} #1 \end{tabular}}
\begingroup
\squeezetable
\begin{table*}[htbp]
\begin{tabular}{p{0.1\linewidth}|P{0.15\linewidth}p{0.3\linewidth}p{0.3\linewidth}}
\multicolumn{1}{c}{\textbf{Item}} &
  \multicolumn{1}{c}{\textbf{Code}} &
  \multicolumn{1}{c}{\textbf{Definition}} &
  \multicolumn{1}{c}{\textbf{Example Student Response}} \\ \midrule
\multicolumn{1}{l}{\multirow{4}{*}{\parbox{0.1\linewidth}{LES1: Why are experiments a common part of physics classes?}}} &
  \centered{\textit{Modeling}} &
  Experiments in class let students develop their own models for phenomena, discover things on their own, and/or develop their own ideas. &
  Because it helps show the process of developing a model, rather than just taking it as fact and using it to solve problems. By studying ``mystery particles'' in lab, we had to experiment and develop our own observations. \\ 
  \cmidrule{2-4} 
\multicolumn{1}{l}{} &
  \centered{\textit{Scientific abilities}} &
  Experiments help cultivate students' scientific abilities, such as experimental design, data collection, and data analysis skills. &
  Experiments provide a way to provide reasoning skills as applied to physics, of which experiments \textit{[sic]} reasoning is needed to have problem [solving] skills not only in the course but in other aspects of life. \\ 
  \cmidrule{2-4} 
\multicolumn{1}{l}{} &
  \centered{\textit{Model testing}} &
  The purpose of doing physics experiments is to prove, support, or test a model.
  &
  Experiments are necessary to test theories. Theories cannot be made into laws without testing. \\ 
  \cmidrule{2-4} 
\multicolumn{1}{l}{} &
  \centered{\textit{Supplemental learning}} &
  Experiments provide supplemental learning experiences for concepts and theories. &
  Experiments are a common part of physics courses because they help you understand the concepts we are learning. \\ \midrule
\multicolumn{1}{l}{\multirow{2}{*}{\parbox{0.1\linewidth}{LES2: Why do scientists do experiments for their research?}}} &
  \centered{\textit{Discoverment}} &
  Experiments contribute to some aspect of the iterative and generative nature of the scientific process aside from testing an existing model. &
  Scientists in the real world are consistently working to provide new findings that deepen our understanding of the world. [There are] plenty of examples in the past, including Newton's laws of motion and evolutionary theory.
  \\ 
  \cmidrule{2-4} 
\multicolumn{1}{l}{} &
  \centered{\textit{Model testing}} &
  The purpose of doing physics experiments is to prove, support, or test a model. &
  You cannot confirm a hypothesis without performing experiments. Without gather repeatable data you cannot decide if something is true or not \\ 
  \bottomrule
\end{tabular}%
\caption{LES items, the codes associated with each, and an example response tagged with each code are shown. The example responses for each code were selected such that each example was assigned only the associated code. Many responses in the data were assigned more than one code. \textit{Scientific abilities} and \textit{Supplemental learning} are drawn from Hu and Zwickl's original codebook \cite{HuZwickl2017}. \textit{Model testing} is equivalent to the original codebook's \textit{Theory testing}. We added the \textit{Modeling} code in response to the recurring presence of its ideas in our dataset and its relevance to our research questions. The original codebook's \textit{Discovery} and \textit{Theory development} codes were merged to create \textit{Discoverment}.}
\label{tab:les}
\end{table*}
\endgroup


Novice responses to these LES items exhibit an almost singular focus on the idea that experiments in instructional labs exist to supplement conceptual learning or test theories (using a layperson's understanding of the term ``theory''). 
The term ``theory'' is somewhat vague here: It has a specific definition in the context of physics, but is used in a lay sense by students, often translating to ``anything that is not an experiment'' or ``what we know from the textbook or lab manual.'' 
Expertlike responses more frequently acknowledge the role of in-class experimentation in the development of scientific abilities and as a means to better understand the scientific process. With regard to experiments in scientific research, novices tend to focus on the notion that experiments exist to test theories. Experts more frequently cite the creation of new models and the iterative nature of experimental model development as purposes of experiments in research.


The original LES \cite{HuZwickl2017} included the codes \textit{Theory testing} (``The purpose of doing physics experiments is to prove a theory or test a hypothesis.'') and \textit{Theory development} (``Experiments inspire the development or improvement of theories.''). 
Due to the vague nature of the term ``theory,'' it is unclear how Hu and Zwickl drew a distinction between ``theory'' and ``hypothesis'' as used by students.
To use the term ``theory'' in our analysis, we would need to establish our own definition, at risk of misrepresenting students' responses in a replication study.

In alignment with ISLE, we remove references to the term ``theory'' in favor of the term ``model.'' A model is a foundational concept in ISLE, used heavily throughout both populations' lab activities. The modified codes we use in our analysis are \textit{\textbf{Model} testing} (``The purpose of doing physics experiments is to prove, support, or test a model.'') and \textit{\textbf{Model} development} (``Experiments inspire the development or improvement of models.''). 

The distinction between \textit{Model development} and \textit{Discovery} (Original definition: ``Experiments help investigate unknowns.'') is subtle and not one we are interested in probing; rather, we are more interested in understanding whether students' responses reflect any acknowledgement at all of steps of the ISLE process aside from model testing. 
Instead, we collapse the two codes into a single \textit{Discoverment} code:
``Experiments contribute to some aspect of the iterative and generative nature of the scientific process aside from testing an existing model.'' 

The final code list is given with definitions and examples in Table \ref{tab:les}.

\subsubsection{Physics self-efficacy}

In context of the physics classroom, self-efficacy refers to students' belief in their ability to practice and succeed at physics. Developing students' self-efficacy is a primary goal for our laboratory instruction, as we want students to walk away from the course with confidence in their ability to design, perform, and interpret experiments \cite{Bandura1977Self-efficacy:Change., Etkina2006ScientificAssessment}. \rem{Physics identity refers to the degree to which students think physics is related to who they are. The literature has shown that physics identity and science identity more broadly are connected to persistence in scientific fields \cite{HazariTheEthnicity}. We seek to understand how NOMR affects each of these constructs.}

This study employs the Physics Identity Survey (PhIS), which we adapted from a science identity survey administered to middle school biology students to evaluate shifts arising from their participation in an immersive virtual lab \cite{Reilly2021AssessingApproach}. The original survey was developed through the lens of Hazari's science identity framework \cite{HazariTheEthnicity}.

The PhIS is divided into two sets of items probing (1) self-efficacy, and (2) physics identity and interest. We focus on the self-efficacy items, listed below, each on the scale [1: Not at all Confident -- 7: Completely Confident]:
\begin{enumerate}[label=\textbf{PhIS\arabic*}]
    \item\label{q:p2} How confident are you that you can design an experiment to answer a scientific question in physics?
    \item\label{q:p3} How confident are you that you can look at data that you collect and characterize its patterns mathematically?
    \item\label{q:p4} How confident are you that you can understand the kinds of problems that experimental physicists would investigate?
    \item\label{q:p5} How confident are you that you could contribute to a team of physicists investigating an experimental physics problem? 
    \item\label{q:p6} How confident are you that you can defend your data analysis to a team of expert physicists?
\end{enumerate}
\rem{We focus on one physics identity and interest item, on the scale [1: Definitely False -- 6: Definitely True]:
\begin{enumerate}[label=\textbf{PhIS\arabic*}]
    \setcounter{enumi}{5}
    \item\label{q:p1} I consider myself a physics person.
\end{enumerate}}

\rem{The physics identity and interest items were adapted from those of the original study by replacing each instance of ``science'' with ``physics''. The interest items gauged students' interest in learning about nature, environmental science, and ecosystems; we referred to students' interest in physics and experimental physics in their place. The self-efficacy} These items were adapted by swapping out learning goals of the ecosystems biology course of the original study for learning goals of the lab courses of concern in this study, e.g., designing an experiment to answer a scientific question in physics and mathematically characterizing patterns observed in data.

We validated the Likert scale items comprising the PhIS in accordance with Adams and Weiman's recommendations for the development of formative assessment instruments \cite{Adams2011DevelopmentThinking}. We conducted think-aloud interviews with eight 100-level physics students. Participants were asked to rate each Likert-scale item and explain their choice as they did so. Participants were recruited through an announcement over the course's web page and incentivized incentivized to participate with \$20 gift cards. The interviews were conducted online, audio recorded, and transcribed by Otter.ai and subsequently hand-corrected.

The interview transcripts were examined to assess the alignment of students' reasoning for the responses they chose with the construct each item was meant to assess. Students' understanding of each item reflected our expectations, and their reasoning for each choice revealed nothing unexpected.
The validation interview results did not lead to any modification of the PhIS.

\subsubsection{\label{sec:flow_background}Flow as a measure of engagement}

We use the psychological theory of flow pioneered by \drc \cite{Csikszentmihalyi1990Flow:Experience} as a lens through which to examine students' engagement with class activities. Known colloquially as being ``in the zone,'' flow is described as a state in which one is completely absorbed in an activity for its own sake, where one action leads smoothly into the next, and one's sense of time becomes distorted. A balance between the person's self-perceived skillfulness at and the challenge posed by an activity is instrumental to achieving a flow state; great challenge must be met with commensurate belief in one's own skill. It is in this state that the most effective learning happens \cite{Csikszentmihalyi2014ApplicationsCsikszentmihalyi.}.

\drc identified seven conditions for a person to achieve flow:
\begin{enumerate}
    \item They know what to do (a clear goal).
    \item They know how to do it.
    \item They are receiving clear and immediate feedback to know how well they are doing.
    \item They know where to go (if navigation is involved).
    \item They see what they are doing as challenging.
    \item They are confident in their ability to complete the task.
    \item Their environment is free of distractions.
\end{enumerate}

As the flow model is fundamentally one of engagement with an activity, it has utility as a measurement of students' engagement with the learning process \cite{Csikszentmihalyi2014ApplicationsCsikszentmihalyi., Rebello2013ProblemFlow, Karelina2022ComparingActivities}. Active engagement is key to learning \cite{Freeman2014ActiveMathematics}; conversely, even well-designed instructional activities with epistemologically authentic inquiry as in ISLE cannot reach students who are not engaged in the learning process.

A subset of the flow conditions comprise an effective basis for maintaining task involvement: A learner needs feedback, confidence in their ability to complete the task, and an environment free of mental distractions. To stick with a task to completion, it is critical a the student's self-efficacy is great enough that they believe they can do so \cite{Bandura1977Self-efficacy:Change.}. To support this belief, they need clear feedback to know how well they are doing and what the next steps are.

Rebello and Zollman note \cite{Rebello2013ProblemFlow} that the zone of proximal development \cite{Zaretskii2009TheDevelopment}, the optimal adaptability corridor \cite{Schwartz2005EfficiencyTransfer}, and flow are all representations of a balance between a learner's skill and challenge. To express the optimal adaptability corridor's dimensions in terms of flow, horizontal transfer (efficiency) maps to skill, and vertical transfer (innovation) maps to challenge. Flow comes in as a means to tie this balance to other affective elements of the student experience, unifying a number of affective constructs in educational psychology under one quantifiable umbrella.




\begin{figure}[htbp]
    \centering
    \includegraphics[width=\linewidth]{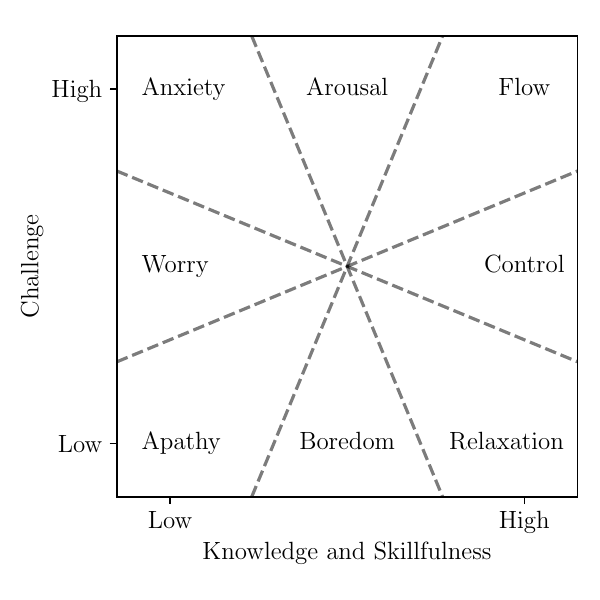}
    \caption{The eight-channel model of flow.}
    \label{fig:flowchannels}
\end{figure}

Massimini and Carli's efforts \cite{Massimini1987TheRehabilitation, Massimini1988OptimalConsciousness.} to develop a quantitative instrument to measure flow led to the eight-channel flow model we use in this study. One begins by constructing a mental state diagram, hereafter referred to as a flow plot, with perceived knowledge and skillfulness on the horizontal axis, and perceived challenge on the vertical axis. Each flow plot is divided into eight channels (Figure \ref{fig:flowchannels}), representing different relative combinations of challenge and skill. The top-right channel, Flow, is the most productive, representing a great challenge met with commensurate skill. Flow's neighbor channels Control and Arousal represent less challenge and less skill than Flow, respectively. Flow, Control, and Arousal are considered productive channels for learning \cite{Csikszentmihalyi2014ApplicationsCsikszentmihalyi.}. The Relaxation channel represents a surplus of skill and dearth of challenge; its mirror channel Anxiety represents an extreme challenge one feels poorly equipped to handle. The least productive states of Worry, Apathy, and Boredom fill out the lower regions of the challenge-skill space, with skill and challenge both insufficient to support productive engagement.

Karelina, Etkina, Bohacek, Vonk, Kagan, Warren, and Brookes \cite{Karelina2022ComparingActivities} used the eight-channel model to compare students' engagement with content-equivalent ISLE-aligned hands-on and video-based labs; we follow much of their methodology in our study of students' engagement with VR labs.

The Flow Survey (FS) is drawn from Karelina \textit{et al.}'s adaptation of a subset of items from the psychometrically-validated Flow State Scale \cite{Jackson1996DevelopmentScale}. The FS uses items from their adaptation with minor wording changes for our experimental context.

It consists of seven 7-point Likert scale items:

\begin{enumerate}[label=\textbf{F\arabic*}]
    \item\label{q:f1} To what extent was the instructor's assistance needed? [1: Not at all -- 7: A lot]
    \item\label{q:f2} To what extent did you know what to do (goal of the task)? [1: Not at all -- 7: A lot]
    \item\label{q:f3} To what extent did you know how to do it? [1: No idea -- 7: Completely]
    \item\label{q:f4}  To what extent did you know how well you were doing? [1: No idea -- 7: Completely]
    \item\label{q:f5} To what extent was the lab challenging? [1: Not at all -- 7: Extremely]
    \item\label{q:f6} To what extent did you feel knowledgeable and skillful during the lab? [1: Not at all -- 7: Extremely]
    \item\label{q:f7} To what extent was the lab fun and interesting? [1: Not at all -- 7: Extremely]
\end{enumerate}

We follow Karelina \textit{et al.}'s analysis methods to create flow plots using two items: \ref{q:f5} as a measure of percieved knowledge and skill on the horizontal axis, and \ref{q:f6} as a measure of challenge on the vertical axis. Students who give a high score to both items are understood to be in a flow state. 

Karelina \textit{et al.}'s study compared average responses along each axis of the flow plot with two-tailed paired t-tests to determine differences in students' perceived skill and challenge between video and hands-on treatment groups. These quantitative comparisons were backed by visual comparisons of which channels students tended to fall into in each treatment group.

We also make use of \ref{q:f7} (``...fun and interesting?'') to characterize the effect of the novelty of VR on students' experience. Weekly measures of this item allow for comparison across lab activities, e.g., comparing hands-on labs to VR labs. 

\section{Methods}

\subsection{\label{sec:nomr_structure}Structure of NOMR labs}

The intervention examined in this study uses VR labs to 
allow students to experience and analyze physical laws in the context of particle interactions that do not exist in nature or on the Internet. We include the constraint that they be consistent with our universe's physics so that students can rely on their extant physics knowledge when reasoning in the VR space. These fictitious physical laws can be construed as hypothetical mathematical variations on Coulomb's Law. A selection of fictitious phenomena are described in more detail in \cite{Canright2020}.

The virtual apparatus is designed such that it does not give perfect answers; experimental uncertainty is still very much present even in the simulation. The ``right answers'' programmed into the simulation are never shared with students or their TAs, such that the only ``right'' answer is the one that students can make the best case for.

In the virtual lab space, \todo{P4 Add figure with screenshot} students can access force and distance measurement tools and a supply of particles (modeled as hard spheres) exhibiting the behavior(s) they are investigating. These particles can be moved around the space freely, as well as be fixed in place individually. To facilitate creating static arrangements of particles, physics can be temporarily paused in the entire space. As there is no copy of the lab manual nor any means by which to record data while in the headset, the operator relies on their group's interaction and record-keeping. Each group of 3-4 students shares one VR headset, with its display mirrored onto a lab computer.

\hl{Multiple instructional practices are in place to combat gendered task division common to inquiry-based physics labs} \cite{Day2016GenderLaboratory, Holmes2022EvaluatingEquity}\hl{. All groups complete a teamwork agreement at the beginning of the term outlining expectations and norms for their interactions in and out of class. The NOMR lab manuals prompt students to take turns using the headset at multiple junctures. Students are encouraged to have each member of the group use the headset to collect data, as a means of obtaining multiple measures from which to determine a central value and uncertainty for each of their measurements.}

The NOMR labs described in this study are used in two instructional contexts: in introductory calculus-based physics, and in a sophomore-level lab for applied physics majors. The first lab encountered, called Charge and Mint, is comprised of two activities. Introductory students complete these components as two separate labs (VR1 and VR2; see full lab titles and schedule in Table \ref{tab:timeline}), and students in the advanced course complete both components in a single lab session (VR1+2).

First, groups design and conduct an experiment to test whether virtual analogues of electrically charged particles follow some re-scaled version of Coulomb's law. This lab serves to familiarize students with the VR environment and doubles as an opportunity to teach (or review) data linearization.

Second, they take qualitative and quantitative data to create an empirical model for the interaction between fictitious \textit{minty} particles, which behave according to an unknown force law. That force law is not included here in an effort to keep it out of print; instead, we present a handful of observations students might make, and leave the specifics of the model to the reader's imagination:
\begin{itemize}
    \item Minty particles repel when they are near each other, and attract when they are far away. A turnaround point where the force is zero exists at a certain separation between particles.
    \item If two minty particles are brought as close as possible to one another and released from rest, they appear to undergo oscillatory motion. Considering the full range of distances achieved in this motion, the range of distances for which the force between the particles is repulsive seems to be shorter than the range of distances for which it is attractive. 
    \item Beyond the repulsive region, no matter how far apart two minty particles are moved, they continue to exert upon each other a substantial attractive force which increases with distance.
\end{itemize}

Charge and Mint is used as a preparatory lab to get students familiar with the virtual learning environment and comfortable with the idea of developing a mathematical model for a completely unknown phenomenon. Once the preparatory lab is complete, students are given a new, more complex phenomenon to explore and model, without any phenomenon-specific scaffolding. This final lab takes two forms: the one-week Exotic Matter Lab for introductory students (lab VR3), and the three-week Manifold Lab (VR3-VR5) for advanced students.

The Exotic Matter Lab's content is identical to the first week of the Manifold Lab: Every group is assigned a different set of fictitious phenomena (referred to as a \textit{scenario}). This phase allows for differentiated instruction as the TA assigns 
scenarios at the start of class based on their impressions of each group's strengths and weaknesses and the nature of the challenge each scenario poses. Each scenario contains up to three distinct types of particles, all visually identical on creation, picked from at random whenever the user creates a new particle. The phenomena underpinning each scenario, and the subject of students' inquiry, are the force laws governing the interactions between the particles. In most cases, a single force law dictates the interaction between each pair of particles, though students may develop different valid interpretations supported by their data. The force laws programmed into NOMR are never shared with students or TAs.

In this model-generating experiment, students are told that they have at most three distinct types of particles in their scenario and given tools to label particles and temporarily remove particles from play.  Their goals are to determine how many types of particles they have, develop a procedure for identifying an unknown particle, and come up with a testable empirical model describing some subset of the behaviors they observe. Students write up their findings in a full lab report. For introductory students, this model-generating experiment report marks the end of their foray into VR.

Advanced students working through the Manifold Lab instead submit reports describing their model-generating experiments and resulting models to a class-wide repository. 
During class in the second week, each group selects another group's report describing a model of a scenario they had not yet interacted with themselves. 
They write and submit a proposal before the third week of lab, describing an experiment to test the other group's model.
These experiments are carried out in the third week, and their results presented in an oral talk symposium in the fourth week.

The Manifold Lab is presented to students with a game-like narrative in which they function as research scientists. 
They explore different ``pocket'' universes with novel forms of matter obeying fundamental force laws unknown to our universe. 
The privilege to conduct the second experiment with better equipment depends on applying (non-competitively) for grant funds: Before performing the second experiment, students write a single-page  ``grant proposal'' in which they summarize another group's findings, propose an experiment to test their model, and request additional or improved equipment within the VR lab. 
The instructor serves as an entity equivalent to the NSF and its reviewers: They review students' grant proposals for feasibility and work with each group to revise proposed experiments such that it is likely that each will produce a clear outcome that builds on the prior group's findings.
Each group receives a few credits to spend on equipment, e.g. more precise measurement tools, a larger workspace, tools that snap to more convenient configurations, or the ability to automatically pause physics after a set amount of time. Occasionally, a clever idea from a group of students inspires the development of a new tool in NOMR, which is added to the upgrade options going forward.

This design seeks to emulate the experience of working within a professional scientific collaboration: The class as a whole collaborates by sharing data and designing experiments to test and revise each other's models. In doing so, students complete an entire cycle of the ISLE process: One group creates a model through a model-generating experiment, another tests it with a testing experiment, and those results serve to reject, revise, or further substantiate the model.

\subsection{Instructional Context}

The study activities took place at the University of Washington in Seattle (UW), a large R1 public research university in the Pacific Northwest. Of the population of students enrolled in the courses examined in this study, 65\% identify as male, and 35\% as female (non-binary gender identities are not reflected in UW records, and we did not solicit this information from students separately). White (41\%) and Asian (36\%) students make up the majority of the population, followed by students who identify with two or more races (8.7\%), Hispanic or Latino,a,e students (7.6\%), and Black students (2.8\%). 

\begin{table}[htbp]
    \centering 
    \includegraphics[width=\linewidth]{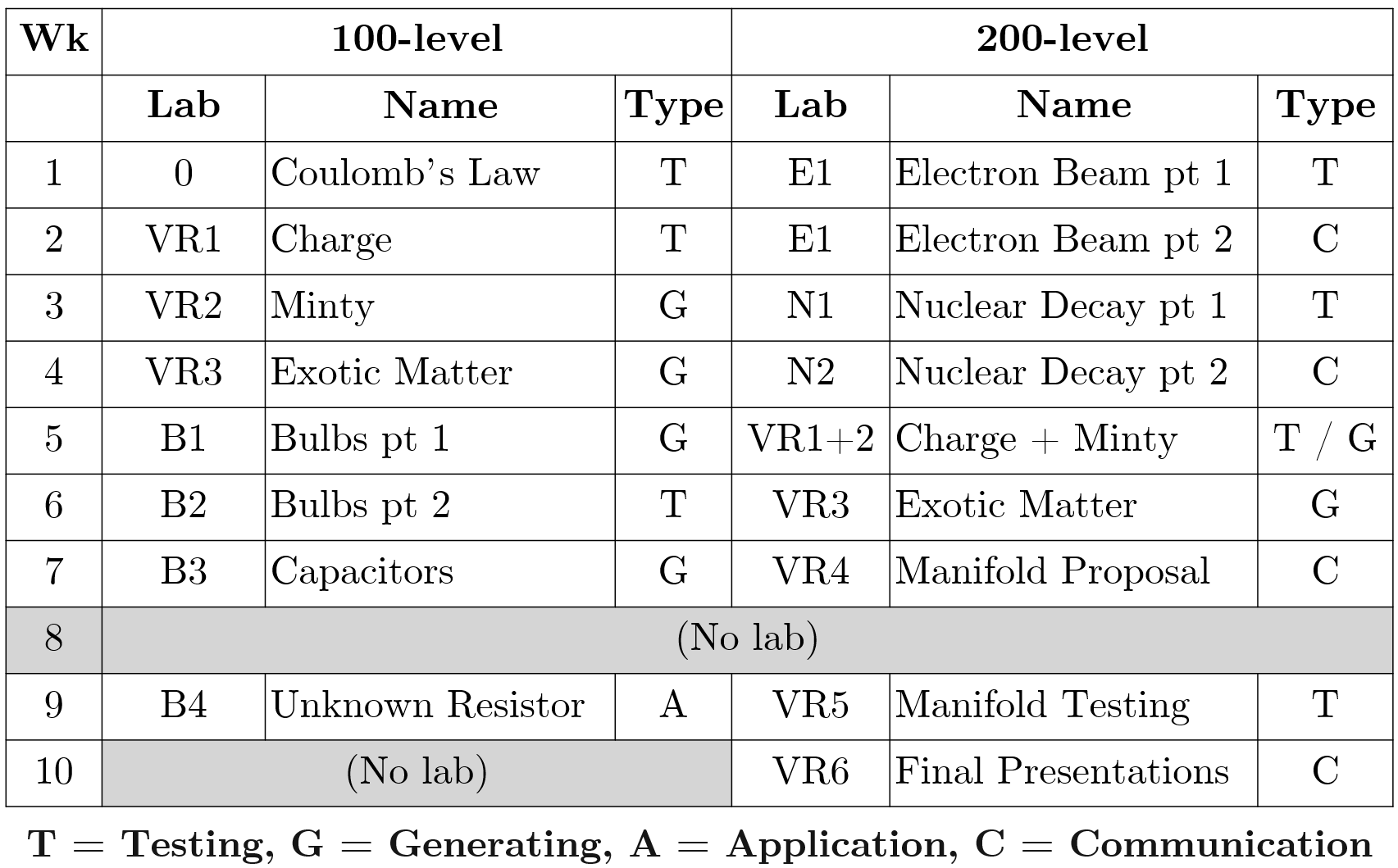}
    \caption{This timeline of events shows the curriculum 100-level and 200-level students completed over the course of the quarter, and when each survey was administered to each population. All LES \& PhIS pre- and post-surveys were administered outside of class time except for the 200-level post-test, which students completed in class after their final presentations. The labels \textbf{G}, \textbf{T}, \textbf{A}, and \textbf{C} represent model \textbf{generation} experiments, model \textbf{testing} experiments, \textbf{application} experiments,  and \textbf{communication}, respectively.}
    \label{tab:timeline}
\end{table}

Our data come from two physics courses at UW during Fall 2022. All instruction was held in person except in the event a student couldn't attend a lab due to illness, in which case their lab partners brought them in via video call, when possible. In both courses, groups of 3-4 students worked together for the entire quarter. Each class's lab curriculum schedule is shown in Table \ref{tab:timeline}.

\begin{description}
\item[100-level population] NOMR was implemented during calculus-based electromagnetism, the second of a three-quarter introductory physics sequence. 467 students enrolled in Fall 2022, and 380 students consented to participate in the study. We refer to the consenting students as the \textit{100-level population} hereafter. This course consisted largely of engineering (74\%) and science (17\%) students filling prerequisites for their major. Students met weekly for three 1-hour lecture sections, a 1-hour tutorial section, and a 2-hour lab section.
\item[200-level population] 
Introduction to Experimental Physics used NOMR as well. This course enrolled 38 students, mostly applied physics majors (55\%) who typically intend to follow an industry-oriented path after graduation, alongside other physics and astronomy majors (21\%), pre-science majors (13\%), computer science majors with physics minors (8\%), and one math major. All 38 students consented to participate in the study; we refer to them as the \textit{200-level population} hereafter. Students met weekly for a 1.5-hour lecture section and 3-hour lab section. \hl{Eight members of the 200-level population had previously seen NOMR labs in the modern 100-level labs; all other 200-level students had taken traditional or online (due to COVID) 100-level labs, without VR.}
\end{description}

\subsubsection{Lab activities}

All lab activities in the 100- and 200-level courses are designed in alignment with the ISLE approach. We say ``in alignment'' because a full implementation of ISLE requires integration of the ISLE process across all components of a course (lecture, lab, etc.), which is not the case at UW.

Every lab activity can be categorized as a hypothesis generating, hypothesis testing, or application experiment (as in Table \ref{tab:timeline}), excluding weeks in the 200-level course dedicated specifically to writing and communication. Students' work is guided and assessed with the ISLE scientific abilities rubrics \cite{Etkina2006ScientificAssessment}.

The first four weeks of the 100-level labs (labs 0, VR1-VR3) focus on particle interactions. Lab 0 is a qualitative testing experiment on Coulomb's law. Students test the effects of charge and separation on the electric force between a copper sphere and a Teflon rod. This is the simplest lab of the quarter, and deliberately so. It is the first lab of the quarter, when students are still joining the course and switching between sections. The following three weeks (labs VR1-VR3) are NOMR labs, as described in section \ref{sec:nomr_structure}.

The remaining four weeks of labs are traditional hands-on labs exploring circuits:
\begin{description}
    \item[B1] Students begin their exploration of circuits with a model-generating experiment seeking to develop a model to describe the behavior of battery-bulb circuits in series, parallel, and mixed configurations.
    \item[B2] Models generated in the prior week are tested against a mixed-configuration circuit. Students create predictions for the current and voltage through each element of the circuit based on their model from the prior week, build the circuit, collect data, and compare the results to their predictions. Where there is disagreement, students revisit and revise their model.
    \item[B3] Capacitors are introduced. Students perform a model-generating experiment to develop a mathematical model for the voltage across a charging capacitor.
    \item[B4] Students are given a resistor of unknown value and a model of voltage across a discharging capacitor. Using this model, students perform an application experiment to determine the value of the resistor, with uncertainty, by manipulating the model to give the resistance in terms of the slope of a linearized plot and the capacitance of the capacitor.
\end{description}

The 200-level course opens with the electron beam lab (E1-E2). Each group is given an electron beam apparatus (commonly called an ``e/m apparatus'') that fires electrons across a user-specified voltage into a helium-filled bulb subject to an approximately uniform magnetic field generated by Helmholtz coils outside the bulb. 
Students are asked to devise and answer a scientific question with the apparatus. Most often, this ends up being a testing experiment based on students' knowledge of electrons' motion in a magnetic field. Occasionally, it turns into a model-generating experiment if a group does not recall this model.

The subsequent nuclear decay lab (N1-N2) works in similar fashion: Students are given an apparatus, instructed in its operation, and are set loose to devise and answer a scientific question of their choosing. In this case, the apparatus is a radioactive Cs source, a Geiger-Muller tube with event counting hardware and software, a box of barriers of various material and thickness, and a stand for all of the above with slots in which to place the source and barriers.

The rest of the 200-level labs are NOMR labs documented in section \ref{sec:nomr_structure}: Charge and Mint (VR1+2) and the Manifold Lab (VR3-VR6).

\subsection{Data Collection}


\subsubsection{\label{sec:methods-lesphis}Lab Epistemology Survey and Physics Identity Survey}




The LES and PhIS were administered as part of the same survey in all cases.

100-level students completed the pre-survey in Week 2, after Lab 0, which was a traditional hands-on lab, and before VR1, the first NOMR lab. The survey was included as part of a timed quiz, and we recognize the time constraint may have influenced students' responses. The post-survey was included as part of an untimed reflection on the performance of their group. It was administered in Week 4 after the conclusion of VR3, the final NOMR lab for 100-level students. Therefore, the pre-post shifts reported here reflect changes in 100-level students' responses before and after only the NOMR labs.

200-level students completed the pre-survey in Week 1 before the start of classes, and completed the post-survey in Week 10, after their final presentations. 
The pre-post shifts reported here reflect changes in students' responses before and after all 9 weeks of labs, including non-NOMR activities.

\subsubsection{\label{sec:flow_data_collection}Flow Survey}

The FS was administered every week at the end of lab to both populations. 
We offered a small amount of extra credit for each week the survey was completed, and emphasized that the surveys would help improve the lab curriculum in future terms. In the 100-level population, 42 students completed all eight surveys; in the 200-level population, 14 students completed all eight surveys. The findings reflect responses only from these students who completed all eight surveys.

\section{Findings}


\subsection{\label{sec:findings_les}RQ1: \rqone}


\begin{figure}
    \centering
    \includegraphics[width=\linewidth]{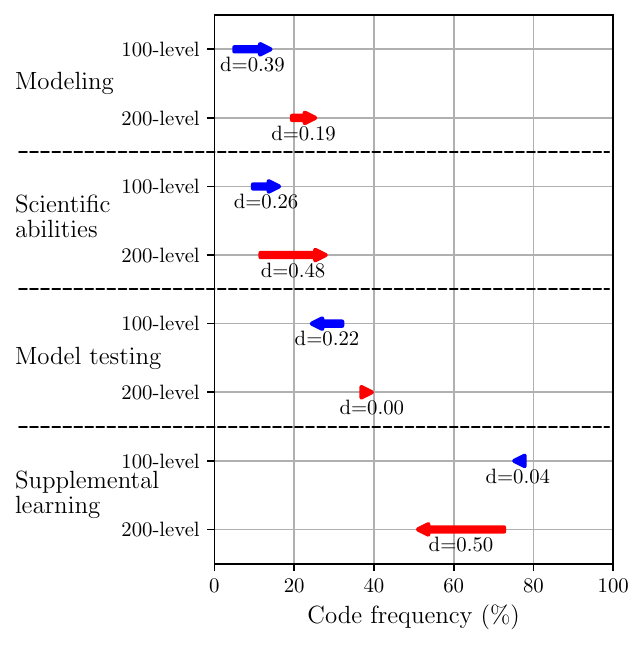}
    \caption{Code frequencies for both populations' responses to \ref{q:1}. A code frequency represents the percentage of responses in a population that were assigned that code. Each response could be assigned zero, one, or multiple codes, so percentages do not add to 100\%. Data from our 100-level population are in blue, and the 200-level population in red; the tail and head of each arrow represent pre-intervention and post-intervention data, respectively. This graphic represents $N_{100}=278$ and $N_{200}=38$ matched pre-post responses from 100-level and 200-level students, respectively. \hl{Cohen's $d$ is calculated according to Eqn.} \ref{eq:1}.
    }
    \label{fig:LES1}
\end{figure}

Responses to \ref{q:1} and \ref{q:2} were coded independently by the first author and another researcher. 
The researchers met to reconcile disagreements after the first coding pass, with a disagreement rate of roughly $10\%$ for \ref{q:1} and $30\%$ for \ref{q:2}. 
After further expanding on the existing code definitions, adding examples, and adjusting the codes as described in Section \ref{sec:les_background}, we reached $>95\%$ inter-rater agreement across all codes. 

The LES findings are presented as code frequencies: the fraction of responses in a population that were assigned each code. Figures \ref{fig:LES1} and \ref{fig:LES2} depict the shift from pre to post for each population-item. Each student response could be assigned no code, one code, or more than one of the codes associated with the item. As most responses were assigned one or more codes, the total number of codes is greater than the number of responses.

\hl{Cohen's $d$ is presented for each shift, calculated from pre and post code frequencies $p_{pre}$ and $p_{post}$:}
\begin{equation}\label{eq:1}
    d=\dfrac{\abs{p_{post}-p_{pre}}}{\sqrt{\frac{\sigma^2_{pre}+\sigma^2_{post}}{2}}};\:\:\: \sigma_i=\sqrt{p_i(1-p_i)}
\end{equation}

\hl{Whether a code is or is not assigned to a given response is a binary variable, so the binary standard deviation is used.}

\subsubsection{LES1: Why are experiments a common part of physics classes?}

Responses to \ref{q:1} were assigned up to four codes, defined with examples in Table \ref{tab:les} and reproduced below:

\begin{description}
\item[Modeling] Experiments in class let students develop their own models for phenomena, discover things on their own, and/or develop their own ideas.
\item[Scientific abilities] Experiments help cultivate students' scientific abilities, such as experimental design, data collection, and data analysis skills.
\item[Model testing] The purpose of doing physics experiments is to prove, support, or test a model.
\item[Supplemental learning] Experiments provide supplemental learning experiences for concepts and theories.
\end{description}

We added the \textit{Modeling} code in response to the recurring presence of its ideas in our dataset and its relevance to our research questions. It captures the creation of new models that results from model-generating experiments in the ISLE process, while \textit{Model testing} represents the subsequent model-testing function of a testing experiment. 

Example responses that were assigned each code are given in Table \ref{tab:les}. A sample response to \ref{q:1} that was assigned multiple codes follows:
\begin{quote}
    ``Experiments allow us to challenge what we know while apply what we have learned. Its a new way of learning things--a more hands-on approach. We can learn about the scientific models and how experiments are designed to either explain a new phenomena or test a pre-existing model.''
\end{quote}

This response is assigned \textit{Modeling} for the phrase ``We can learn\ldots how experiments are designed to either explain a new phenomena\ldots'' and \textit{Model Testing} for the last part of that sentence: ``\ldots or test a pre-existing model.'' \textit{Supplemental learning} is present in a couple of places: ``We can learn about the scientific models\ldots'' and ``Its [sic] a new way of learning things--a more hands-on approach.'' 
Finally, the clause ``\ldots how experiments are designed\ldots'' merits the \textit{Scientific abilities} code.


Both populations' code frequencies are plotted in Figure \ref{fig:LES1}. Comparing pre-post results, we find that the students in our study became more likely to indicate a belief that labs are meant to develop their \textit{Scientific abilities} and give them opportunities to develop their own models (\textit{Modeling}). The 100-level population became less likely to cite  \textit{Model testing} as a purpose of in-class experiments, while the 200-level population became dramatically less likely to cite \textit{Supplemental learning} for the same.

\subsubsection{LES2: Why do scientists do experiments in professional research?}

\begin{figure}
    \centering
    \includegraphics[width=\linewidth]{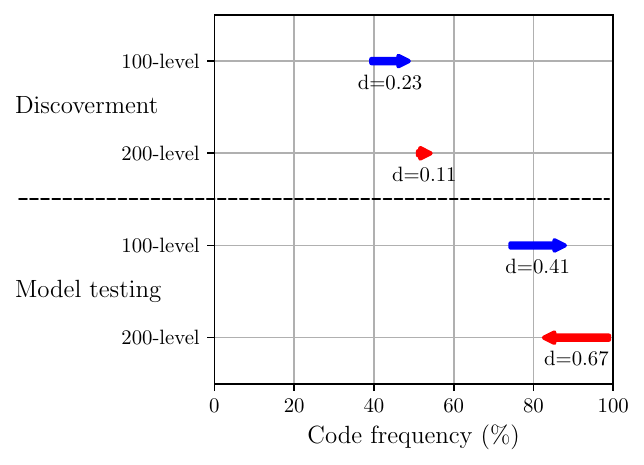}
    \caption{Code frequencies for both populations' responses to \ref{q:2}. A code frequency represents the percentage of responses in a population that were assigned that code. Each response could be assigned zero, one, or multiple codes, so percentages do not add to 100\%. Data from our 100-level population are in blue, and the 200-level population in red; the tail and head of each arrow represent pre-intervention and post-intervention data, respectively. This graphic represents $N_{100}=278$ and $N_{200}=38$ matched pre-post responses from 100-level and 200-level students, respectively.
    \hl{Cohen's $d$ is calculated according to Eqn.} \ref{eq:1}.
    }
    \label{fig:LES2}
\end{figure}



Responses to \ref{q:2} were tagged with up to two codes, defined with examples in Table \ref{tab:les} and reproduced below:
\begin{description}
\item[Discoverment] Experiments contribute to some aspect of the iterative and generative nature of the scientific process aside from testing an existing model.
\item[Model testing] The purpose of doing physics experiments is to prove, support, or test a model.
\end{description}

\ref{q:2}'s code frequencies are plotted in Figure \ref{fig:LES2}. Both populations acknowledged the iterative and generative elements of the scientific process (\textit{Discoverment}) more frequently after NOMR labs than at the beginning of the course. The 100-level population's \textit{Discoverment} code frequency increased by a significant degree; the 200-level population's frequency started higher and saw a smaller increase. \hl{The 200-level change is small enough to be statistically insignificant.} 
\rem{Roughly 20\% of the 200-level students had also experienced NOMR labs at the 100-level in a prior course.}


The 100-level population was assigned \textit{Model testing} codes more frequently after NOMR labs than at the beginning of the course, jumping from 73\% to 89\%. 
The 200-level population's code frequencies saw a significant decrease in \textit{Model testing}, moving from 100\% to 82\% of matched responses.





\subsection{RQ2: \rqtwo}

\subsubsection{Statistical Analysis}

\todo{P2 Many citations still to add here.}

We assess pre-post shifts in students' responses to the Likert items comprising the PhIS by calculating 
$p$-values using the Wilcoxon signed-rank test \cite{Wilcoxon1945IndividualMethods}. We chose the Wilcoxon test because it is a nonparametric test and thus makes no assumptions about whether the dataset is normally distributed. 
We opted against using the Mann-Whitney U test as we are interested in testing for differences within paired samples. The Wilcoxon signed-rank test is a more appropriate choice than the Mann-Whitney U test, since the Mann-Whitney test is used for independent samples. \todo{P2 cite}

\begin{figure*}[htbp]
    \centering
    \includegraphics[width=\linewidth]{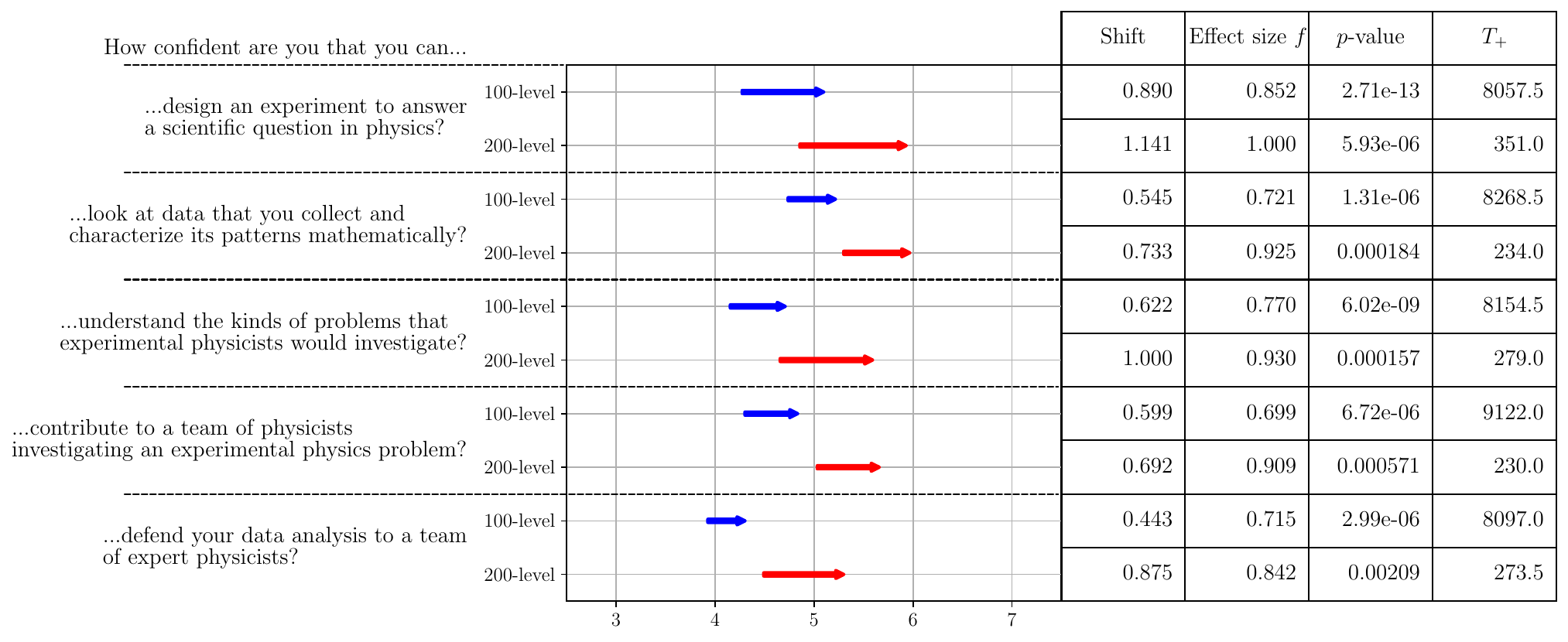}
    \caption{PhIS self-efficacy findings for all students who completed both the pre- and post-surveys. 100-level data ($N_{100}=225$) are in blue, 200-level data ($N_{200}=35$) in red. Each arrow represents the difference between the interpolated median of the pre-survey (tail) and post-survey (head) responses for the associated population-item. 
    }
    \label{fig:phis}
\end{figure*}
\begin{figure*}[htbp]
    \centering
    \includegraphics[width=\linewidth]{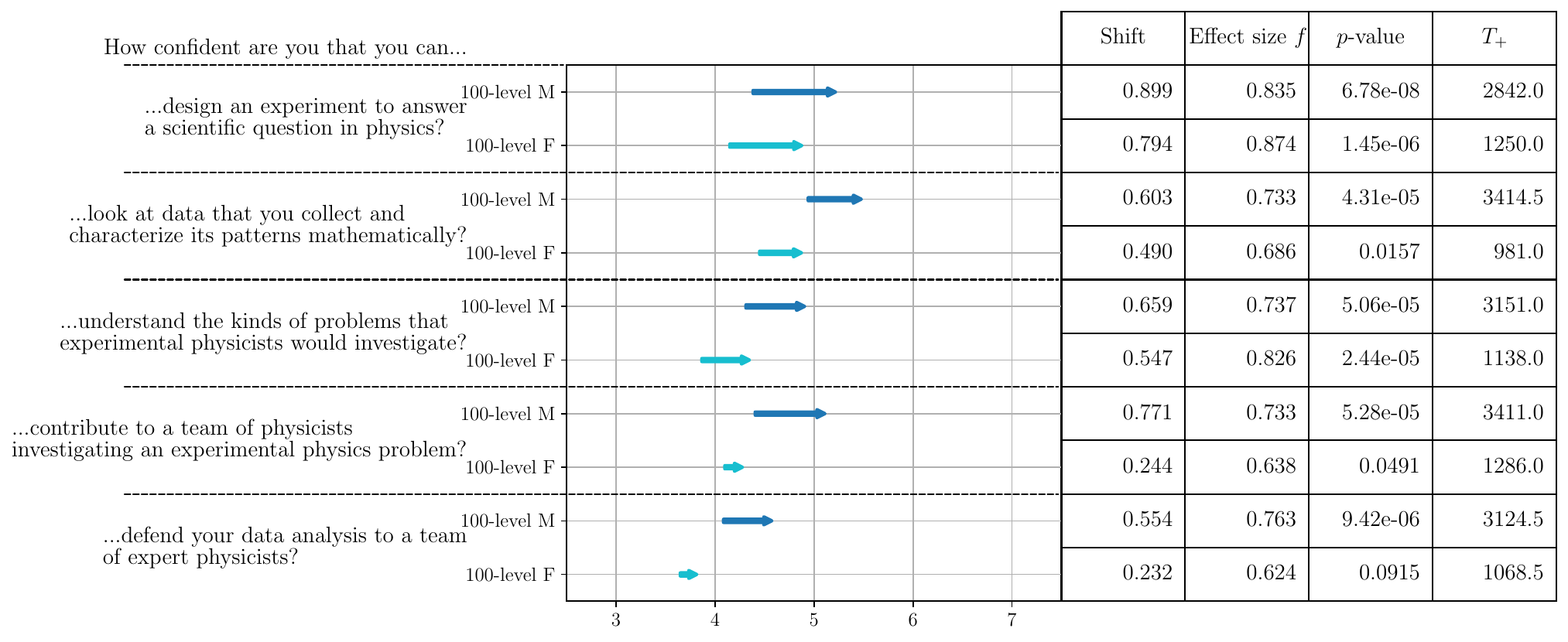}
    \caption{PhIS self-efficacy findings for 100-level students who completed both the pre- and post-surveys, divided by gender. Data from male-identifying students ($N_{M}=135$) are in dark blue, female-identifying students ($N_{F}=88$) in light blue. Each arrow represents the difference between the interpolated median of the pre-survey (tail) and post-survey (head) responses for the associated population-item. 
    }
    \label{fig:phisgen}
\end{figure*}

We report the common-language effect size $f$ computed from the Wilcoxon test statistic $T$ and the total rank sum $S$ by the expression: \todo{P1 cite} $f=\frac{1+T/S}{2}$.  
In general terms, $f$ tells us what fraction of students reported a higher score in the post-survey than in the pre-survey. However, this metric does not account for ties, where a student gives the same response for an item in the pre- and post-surveys; ties are ignored in the calculation of the effect size. The common-language effect size can range from 0 to 1 for a given item, where 0 indicates that all respondents gave an equal or lower score to the item on the post-survey than on the pre-survey, 0.5 indicates that as many respondents reported a higher score as reported a lower one, and 1 indicates that every respondent's post-survey score was equal to or higher than their pre-survey score.

\subsubsection{Self-efficacy}

As shown in Figure \ref{fig:phis}, positive shifts are observed for both populations in all self-efficacy items. 
The figure shows interpolated medians for each item before and after the intervention; each arrow's tail represents the pre-intervention median, and each head the post-intervention median. Thus, the length represents pre-post change. The 100-level and 200-level populations' data are shown in blue and red, respectively. The effect sizes vary, but there is a positive shift for every self-efficacy item at a 99.7\% confidence level or better ($p<0.003$) in both populations.

Comparing the size of the populations' arrows for each item, we note that the shift in the 200-level students' responses is consistently greater than that of the 100-level students, by roughly 50\% on average. This magnified effect is observed despite the 200-level pre-data medians being consistently higher than those of the 100-level students, leaving less room for improvement.

Every individual 200-level response to the item ``How confident are you that you can design an experiment to answer a scientific question in physics?'' either did not change or became more expertlike. 
This is a useful example for understanding the common-language effect size. Of all the students whose scores changed, 100\% of them increased. Therefore, $f=1$. 

\hl{The PhIS results are unique among our dataset in that there are notable differences between responses given by male- and female-identifying students, as shown in Figure} \ref{fig:phisgen}. \hl{While there are positive shifts for all items for both genders, female-identifying students consistently started lower on each item and saw a smaller increase. Female-identifying students' increases on the last two items are notably small; the shift in responses to} \ref{q:p5} \hl{barely meets the traditional significance threshold $p<0.05$, and the shift for} \ref{q:p6} \hl{does not.}


\rem{Physics identity item \ref{q:p6} (``I see myself as a physics person.'') has featured prominently in prior research \cite{HazariTheEthnicity}. \todo{P3 find 1-2 more} 
We observed very small shifts in both populations,
neither of them statistically significant. }
\todo{P3 add phys identity findings to appendix}



\subsection{RQ3: \rqthree}

\begin{figure*}[!htbp]
    \centering
    \includegraphics[width=\linewidth]{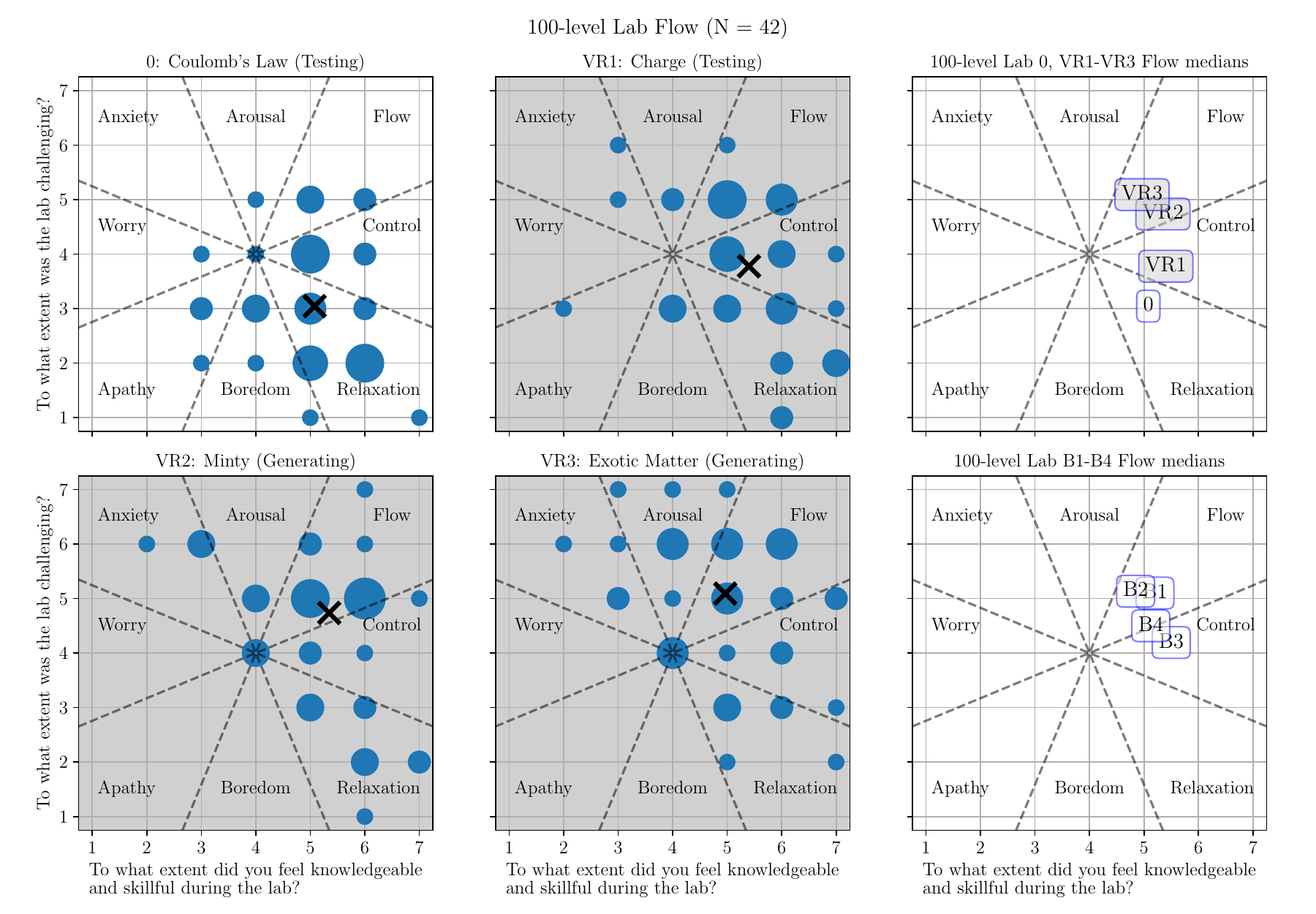}
    \caption{Flow plots for all $N_{100}=42$ 100-level students who completed all eight flow surveys. Shaded plots represent VR labs. The black cross on each plot represents the 2D interpolated median of those responses. The area of each dot is proportional to the number of students it represents. The upper right and lower right plots show the medians from the first four and last four labs of the quarter, respectively.}
    \label{fig:flow}
\end{figure*}

\begin{figure*}[!htbp]
    \centering
    \includegraphics[width=\linewidth]{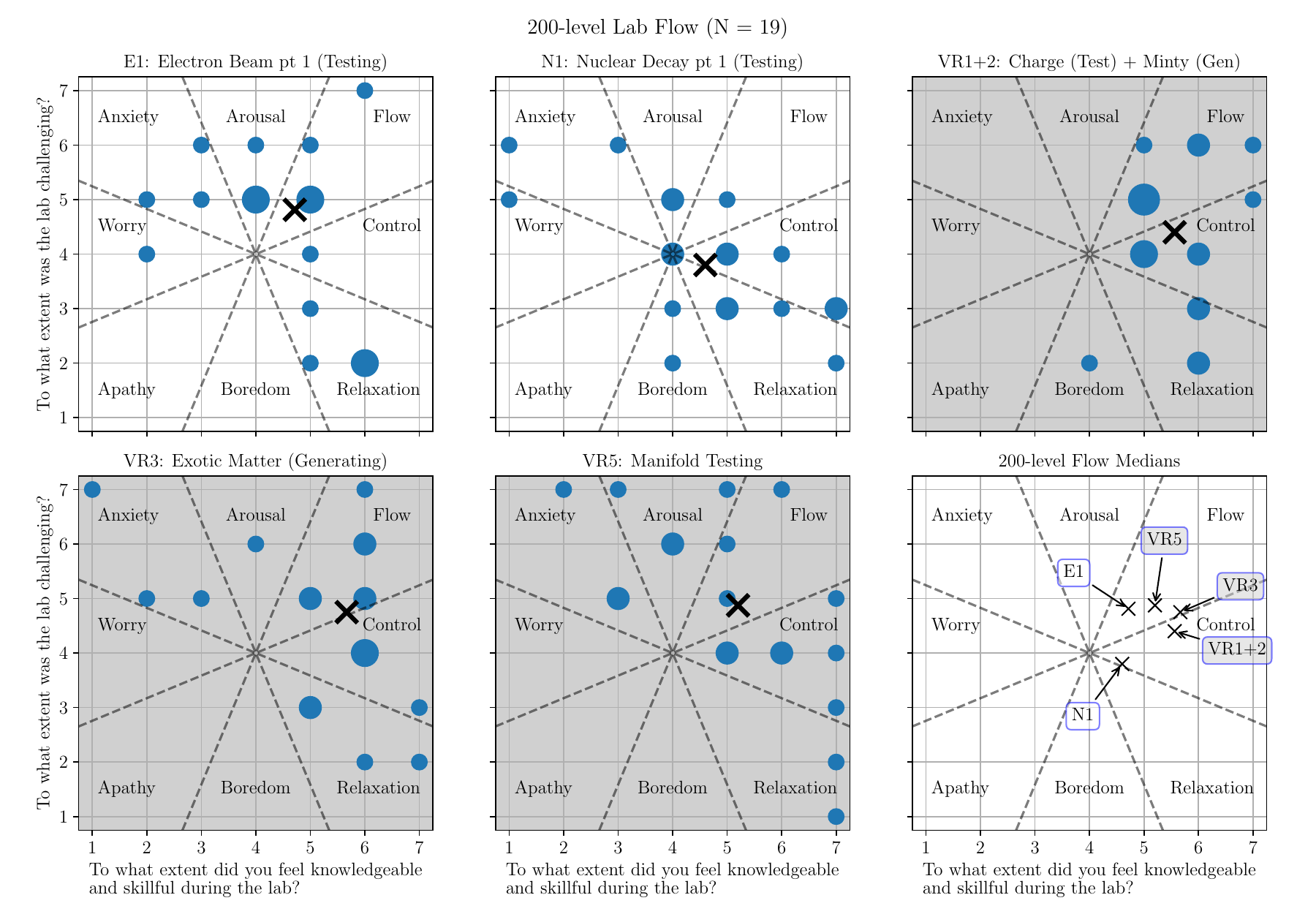}
    \caption{Flow plots for all $N_{200}=19$ 200-level students who completed every flow survey. The sixth (lower right) plot shows the medians from the other five plots all together. Shaded plots and labels represent VR labs. The black cross on each flow plot represents the interpolated median of those responses. The area of each dot is proportional to the number of students it represents. Lab activities focusing on communication and writing with no new experimentation, labeled ``C'' in Table \ref{tab:timeline}, are omitted.}
    \label{fig:flow2}
\end{figure*}

Following the analysis methods of Karelina \etal \cite{Karelina2022ComparingActivities} described in Section \ref{sec:flow_background}, we produced flow plots for each week of each population's lab activities. 
The number of students who responded with each $(x,y)$ pair, representing (skill/knowledge, challenge) is indicated by the size of the dot at that point. The \textit{area} of each dot is proportional to the number of students it represents. For example, suppose 4 students responded that the lab was extremely challenging ($y=7$) and they felt only moderately skillful ($x=4$), and 16 students responded that the lab was a significant challenge ($y=5$) that they felt prepared to tackle ($x=5$). We would see a dot at $(4,7)$ with radius $R$ and a second dot at $(5,5)$ with radius $2R$. The absence of a dot indicates that zero students gave the associated response.

For our analysis, we invent a quantity to help characterize the quality of student engagement through the lens of flow across an entire class. We calculate the interpolated median of these data along each dimension of the flow plot (skill/knowledge on the $x$-axis, challenge on the $y$-axis) and plot the point defined by these medians. As a metric for the aggregate engagement state of the class, the closer to the top-right corner this 2-D median is, the more effective the lab was at inducing productive engagement. We use the interpolated median rather than the mean, as it better captures the distribution of these ordinal data. 

\hl{One can produce error bars for the interpolated medians by taking the standard error along each dimension. We found this to consistently produce error bars the same size or smaller than the marker, so we have omitted them.}

The 100-level population's flow data from labs 0-VR3 (the first half of the course) are shown in Figure \ref{fig:flow} alongside summaries of the 2D medians for labs 0-VR3 and B1-B4.  
The 200-level population's flow data are plotted and their medians summarized in Figure \ref{fig:flow2}. 



We note a few key findings from the flow plots of 100-level students (Figure \ref{fig:flow}):
\begin{enumerate}
    \item  In both 100-level model-generating VR labs (VR2 and VR3), zero respondents reported Boredom, Worry, or Apathy. This is not the case in any other 100-level labs.
    \item There is gradual migration out of Relaxation from labs 0-VR3.
    \item The especially high-challenge, high-skill point in the Flow channel $(6,6)$ had zero respondents in the first two labs 0 and VR1, one respondent in lab VR2, and several respondents in lab VR3.
\end{enumerate}

\begingroup

Looking at the migration in the 2D medians of the 100-level students' responses to labs 0-VR3 (Figure \ref{fig:flow}, top right), we see a trend consistent with the design of the curriculum: it starts out easy and becomes more difficult by the week (upward movement), but students report feeling equipped to handle the increased difficulty (remain on the right side). Lab VR3, the third and final week of NOMR lab activities, had the median closest to the top-right of the plot out of all of the VR labs. 

The rightmost column of Figure \ref{fig:flow} lets us compare the 2-D medians of the NOMR labs to the ISLE-based hands-on circuit labs B1-B4, shown on the top-right and bottom-right of the figure, respectively.
The students complete all A labs before embarking on the B labs, which comprise the latter half of the course. We note that lab VR3 and the first two weeks of circuit labs (B1 and B2) all achieved similar states of productive engagement, landing near the diagonal in the Flow channel. 


In the 200-level population, $N_{200}=19$ students completed every survey over the course of the quarter; the flow plots and aggregated medians are shown in Figure \ref{fig:flow2}. Lab activities focusing on communication and writing with no new experimentation (Table \ref{tab:timeline}), are omitted from the flow plots. As with the 100-level population, the VR labs' medians follow a path up and to the left over time, starting in Control and landing solidly in Flow. The median corresponding to the final VR lab (VR5) is farthest along the diagonal bisecting the Flow channel out of all the labs. 
We note that only 3 student responses for VR5 are actually in the Flow region, with the majority sitting in Control and Arousal and a few in Anxiety and Relaxation. The Charge and Mint lab (VR1+2) and Exotic Matter lab (VR3) had the most individual responses in the Flow region at 7 out of 19.

\begin{figure}
    \centering
    \includegraphics[width=\linewidth]{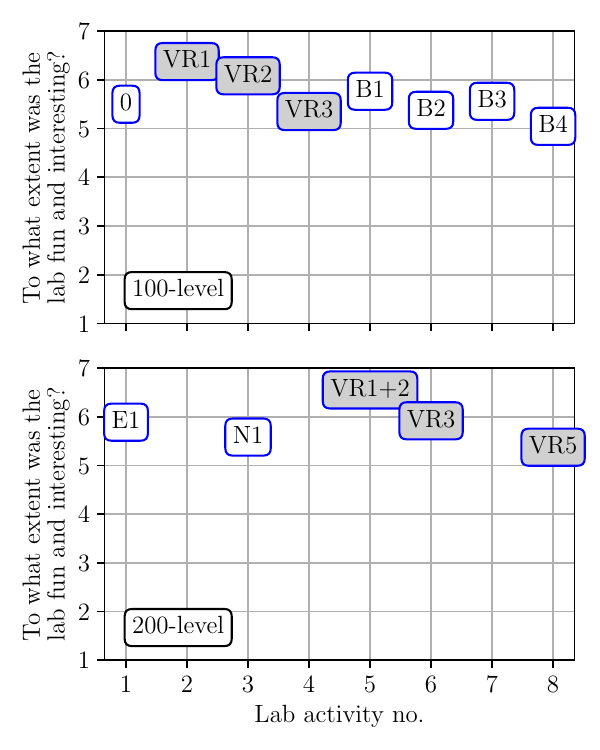}
    \caption{The top and bottom plots show the 100-level ($N_{100}=42$) and 200-level ($N_{200}=19$) populations' interpolated median response each week to \ref{q:f7} (``To what extent was the lab fun and interesting?''), respectively. NOMR labs are shaded. Lab activities focusing on peer review and writing in the 200-level class are omitted.}
    \label{fig:flowfun}
\end{figure}


Figure \ref{fig:flowfun} shows both populations' responses to \ref{q:f7} (``To what extent was the lab fun and interesting?'') for each lab activity. At the 100-level, students' responses
are very high for VR1, remain elevated for VR2, and their responses for VR3 are indistinguishable from any other activity.

We see a similar pattern in the 200-level students' responses to the three VR labs in their curriculum: The Charge and Mint lab (VR1+2) was seen as the most fun (median $\sim6.7/7$), followed by the subsequent Manifold lab part A (VR3) with a half-point lower median, and the final Manifold lab part C (VR5) landed between the hands-on labs with a median of $\sim5.5/7$.

Both populations reported the highest flow state in the last week of NOMR activities. The last NOMR lab for 100-level students was a model generating experiment (VR3). For 200-level students, it was a test of models generated in a prior experiment (VR5).

\par
\endgroup


\section{Discussion}

\subsection{RQ1: \rqone}

\subsubsection{LES1: Why are experiments a common part of physics classes?}

The data show (Figure \ref{fig:LES1}) that for both populations in our study, the frequency of each code underwent a shift consistent with growth toward an expertlike understanding of the role of experimentation in physics classes. 
\hl{Both populations started with the \textit{Modeling} code occurring much less frequently than \textit{Model testing}. After the intervention, both received the \textit{Modeling} code more frequently and the 100-level population received \textit{Model testing} less frequently than when they started. In particular, the 200-level population closed the gap entirely: Their post-survey frequencies for \textit{Modeling} and \textit{Model testing} had equal values at the end of the quarter. }
Movement toward equal emphasis on these two codes is considered expertlike as it is in alignment with the epistemological basis of the ISLE approach, itself grounded in the real-world practice of physics.

\textit{Supplemental learning} came up much more frequently than \textit{Scientific abilities} in pre- and post-surveys in both populations. However, both populations narrowed that gap: \hl{Both populations reported \textit{Scientific abilities} more often and the 200-level population reported \textit{Supplemental learning} less often than when they began, which are shifts to more expertlike belief.}

The small 100-level \textit{Supplemental learning} shift is a little surprising in context of the model-generating NOMR labs, in which students are coming up with models for completely fictitious phenomena in a clear and deliberate departure from lecture content. \textit{Supplemental learning} is not a goal of those labs. It could be that the high and unaffected \textit{Supplemental learning} code frequency is due to students' long history of traditional science labs where supplemental learning is indeed the primary goal; years of conditioning are not readily overcome by a 3-4 week intervention.


In sum, The LES code frequencies in both populations reveal that the students' beliefs are changed by the NOMR labs in significant ways. Regarding experimentation as part of their course-taking, they are less likely to consider classroom laboratory activities to be a supplement to the theory they learn in lecture (primarily in the form of testing theories they have already learned), and more likely to see them as an opportunity to gain new knowledge in the form of developing scientific abilities and developing scientific models. 

\subsubsection{LES2: Why do scientists do experiments in professional research?}

We compare our \textit{Discoverment} code frequencies to the original study's findings by adding together each of their populations' \textit{Model Development} and \textit{Discovery} code frequencies. We expect that a subset of responses in each population received both codes, so these combined \textit{Discoverment} code frequencies are likely over-estimates.


We would expect a collection of expert responses to \ref{q:2} to receive both \textit{Discoverment} and \textit{Model testing} codes at a fairly high frequency. The original study's PhD student responses merited \textit{Discoverment} at 65\% frequency and \textit{Model testing} at 90\% \cite{HuZwickl2017}. These were the greatest and smallest frequencies among all populations in the original study, respectively. PhD students were the most experienced population in the original study, suggesting that expertlike change would manifest as an increase in the \textit{Discoverment} code frequency and a decrease in the \textit{Model testing} code frequency.


The increase in \textit{Discoverment} frequencies in both UW populations suggest growth toward an expertlike understanding of the multi-faceted role of experimentation in scientific research. On account of the low starting point for both populations, we suggest that we can interpret any increase as developing more toward expertlike beliefs.

The \textit{Model testing} shifts are more ambiguous: the increase in 100-level responses is opposite to what we would consider an expertlike change. However, that increase brings the 100-level \textit{Model testing} frequency almost exactly in line with that of the PhD student population in the original study. \hl{On its own, this increase could be generously interpreted to suggest that 100-level students' belief that experiments in scientific research serve to test, support, or prove models was unchanged given the inherent uncertainty in qualitative analysis. More conservatively, this incomplete but accurate belief was bolstered alongside beliefs (i.e. \textit{Discoverment}) that lead to a more complete expertlike understanding.}
The 200-level population saw a decrease in \textit{Model testing} codes, moving from 100\% to 82\% frequency. This is below that of any population in the original study. Unlike the 100-level activities, the 200-level activities--especially the Manifold Lab--are routinely connected to examples of real-world research as part of the introduction for each lab. Explicitly drawing these parallels may have contributed to the relatively large shift in the 200-level responses.


Taken as a whole, these findings
represent growth toward an expertlike understanding of the role of experimentation in physics. After NOMR, students shift away from viewing experimental physics exclusively as a theory-testing endeavor, to one that includes a variety of important aspects of the role of experimentation in generating new knowledge. This shift brings students closer to to the expert view of scientific knowledge as a process that involves rigorous validation in the natural world.

\subsection{RQ2: \rqtwo}

The presence of a positive change at a 99.7\% confidence level or better ($p<0.003$) for every PhIS self-efficacy item for both populations suggests that students' self-efficacy around conducting physics experiments is tangibly improved after participating in NOMR labs. Students believe that they are learning in ways consistent with widely agreed-upon undergraduate physics laboratory learning goals \cite{AAPTLabGuidelines}. That these shifts are observed in the 200-level population is not especially surprising, as each item represents a core learning outcome for a quarter-long course for physics majors that specifically focuses on experimental physics. It is surprising that we see similar shifts in the 100-level population after just a few weeks of NOMR laboratory exercises. Still, the 100-level shifts' lesser magnitude is consistent with the relatively light depth and duration of the 100-level intervention compared with the 200-level version. All told, students' responses moved closer to those of expert physicists and indicate that their confidence in their own ability to do experimental physics is strengthened significantly.

\hl{Looking at the differences by gender in the 100-level data, the lower starting point for female-identifying students across all items is consistent with research into identity and belongingness in introductory physics} \cite{HazariTheEthnicity}. \hl{Put colloquially, the field of physics is commonly thought of as being an old boys' club, and female-identifying students have on average a harder time developing science identity in physics. }

\hl{The dramatically smaller shifts in female-identifying students' responses to} \ref{q:p5} and \ref{q:p6} \hl{are of particular interest. These items are focused on one's extrinsic interactions with a group of physicists than on one's intrinsic ability to perform a category of tasks. For that reason, it is plausible to believe that these items are inherently gendered; that is, administering these two items would elicit a similar difference by gender in any context. It may be the case that male-identifying students build more confidence in NOMR labs than female-identifying students do, but the absence of gender-distinct results in the LES and flow data suggest the interaction with the headsets seems to be gender neutral. Thus, we hesitate to attribute the results from these items to the instrument or the intervention, and highlight this as an area for future study. }


\rem{The physics identity and interest item yielded no measurable pre-post change, which we interpret as evidence that NOMR has done no harm in physics identity formation. In other words, we have not scared students off; in context of an introductory physics course, we interpret that as a positive result.}

Both populations' self-efficacy about designing, conducting, and interpreting experiments is significantly improved after working through NOMR labs. These shifts are aligned with the AAPT laboratory learning objectives \cite{AAPTLabGuidelines}. We suggest these data indicate that the NOMR labs are helping students develop confidence in their professional capacity as experimentalists, while also helping them develop more expertlike habits of mind about experimental physics\rem{, all at no expense to their interest in and identity with physics}.

\subsection{RQ3: \rqthree}


We see the majority of students in the productive zones of Flow, Arousal, and Control during the model-generating NOMR labs.
We interpret being in a flow state as optimized student engagement in the learning activities.
Achieving flow requires that students know what to do, how to do it, and how well they are doing; students tuning out or becoming lost in the face of the open-endedness of the activities would be reflected in low-skill responses on the left of the flow plots. The data show that the NOMR labs are providing just enough scaffolding to keep students in the zone of proximal development and in flow \cite{Rebello2013ProblemFlow}.
Of the first series of 100-level labs (0-VR3), VR3 induced the most productive aggregate state of engagement in students; no responses indicated a state of Worry, Apathy, or Boredom.
The 200-level population achieved more productive engagement with the VR labs than either of the hands-on labs E1 and N1.

We recognize that VR is an engaging environment on its own.
While it may be hard to disentangle the novelty of VR from the activities themselves, we do see evidence that the novelty wears off.
We consider the effect of the gaming/entertainment appeal of VR by examining student responses to item \ref{q:f7} (plotted in Figure \ref{fig:flowfun}): ``To what extent was the lab fun and interesting?'' The novelty effect is associated with the introduction of an exciting new technology in the classroom, which induces an initial boost in student engagement that eventually wanes \cite{Clark1983ReconsideringMedia}.
 
We estimate that VR's novelty lasts two weeks in our context, as we observe in both populations the highest \ref{q:f7} score in the first NOMR lab, the second highest in the second NOMR lab, and the third NOMR lab is no more or less fun than any of the hands-on labs in the course.

Despite the third week of NOMR labs not benefiting from the novelty effect, students reported the greatest aggregate flow state during that activity (VR3 and VR5 at the 100-level and 200-level, respectively). This suggests that the novelty effect does not fully explain the
productive and deep engagement with VR physics labs. Students are not engaged simply because VR is fun; they are engaged because the physics is compelling. NOMR labs use VR specifically because its ``secret sauce'' of hands-on interaction with fictitious physical phenomena is otherwise impossible. We consider students' strong engagement after the novelty has worn off as evidence that NOMR labs may be leveraging the unique affordances of VR in a pedagogically useful way.


Our comparison of VR and hands-on labs contrasts with Karelina \textit{et al.}'s comparison \cite{Karelina2022ComparingActivities} between students' engagement with video labs and hands-on labs. They found students reported video labs to be slightly more challenging, less fun, and that they felt less skillful when compared to hands-on labs. Our findings demonstrate that students' engagement with VR labs can be similar to or better than their engagement with hands-on labs in the same course. 
It is important to note that Karelina \etal compared two distinct populations of students who went through hands-on and video versions of the same lab activity, while our study compares responses to different lab activities from the same population, so comparisons between our findings and theirs should be made with caution.

We hesitate to overinterpret our analysis of the flow data. In this study we analyze absolute rather than relative scores. The original application of the eight-channel flow model \cite{Massimini1988OptimalConsciousness.} collected responses from each participant at many points in time over several days. The researchers determined each participant's average response for an item. When creating flow plots, they plotted z-scores relative to each participant's average response to each item. This method accounts for the fact that every individual interprets Likert scale questions differently; one person's 3/7 is another's 6/7. Karelina \etal \cite{Karelina2022ComparingActivities} adapted this original methodology in favor of examining absolute scores due to the limitations of a classroom setting. They had no more than 2-3 responses from any given participant, and we replicated their methodology for comparability.

\hl{Further, we note that flow states manifest in neurodiverse learners in ways that are not fully understood }\cite{McDonnell2014GoingStates}. \hl{The fact that the measurements of students' flow state takes place in a group learning context adds the complexities of group social interactions to the picture. Flow is an individual measurement of an experience that occurs in a group context, which does not give us any information about group dynamics and cohesion. These shortcomings are excellent areas for further work.}

\subsection{Use of fictitious physics with virtual reality}

Existing survey data does not fully capture the in-class and in-headset experience of NOMR labs. 
Students' experimental results in NOMR labs are different from reexaminations of well-understood phenomena: 
Every group's findings are new knowledge students have generated from scratch.
We posit that, in this way, NOMR labs may allow students to experience the satisfaction of discovery that professional physicists find so compelling, as reflected in feedback from post-course surveys:
\begin{quote}
``I was thrilled and enlightened to be put in a position to analyze physical phenomena that were undocumented and that I had never heard of. Being able to work with a pocket universe and using experimentation to describe it was the best experience in physics courses I've ever experienced. I preferred this VR experience to any physical lab for the sole reason of it being entirely new and having to get every ounce of info about it through experimentation and collaboration.''
\end{quote}
\begin{quote}
``It's been so much fun
learning physics in an exploratory way that focuses on letting us be creative with our thinking. I've
not only learned a lot about error analysis and creating models, but also gained a much better
perspective on how science and research `work' in the real world.''
\end{quote}
As NOMR lab instructors and experienced teachers, we observe students engaging in scientific creativity in ways we have not seen before. 
Rigorous characterization of creativity is a challenging research task, and is not captured in the surveys we administered. We postulate that the use of VR is in part responsible for helping unleash student creativity, and highlight this as an area for further study.




This study does not create evidence one way or another about whether VR is necessary for the implementation of the fictitious physics approach to stimulate creativity in introductory labs.
We predict that such an intervention would not be as effective without VR, in that the joy of discovery expressed above would likely be altered by the difficulties associated with non-immersive simulations. Manipulating objects in 3D space on a 2D screen can be challenging, not to mention that it risks widening the conceptual gap between the novel physics and the physics of our universe. Being overly disconnected from a physically interactive 3D space may preclude the suspension of disbelief that allows students to engage so readily in NOMR. It is our experience that the preceding argument is difficult to make convincingly without sharing the in-headset experience; making it rigorously will require substantial human-computer interaction research. 

\hl{There remains a possibility that a non-VR version's effectiveness could be sufficient for a headset-free version of NOMR to be a favorable trade-off, considering the expense and overhead associated with VR technology. Presently, Meta Quest 2 headsets (the same used for this study) cost $\$300$ each; at eight lab groups to a section, the cost of outfitting a classroom to run NOMR labs is $\sim\$2400$. While not outrageous, this is beyond the reach of many physics laboratory budgets, especially in under-served communities. Phone-based virtual or augmented reality, 3D simulations experienced on a monitor and controlled via mouse and keyboard, or entirely 2D simulations could all employ fictitious physics in a similar way to NOMR labs without the expense of full VR headsets. Further developments based on the work in this paper can contribute to exploring these avenues to open up broader access to this instructional innovation.}


\begingroup
\squeezetable
\begin{table*}[htbp]
\begin{tabular}{p{0.32\linewidth}>{\centering}m{0.05\linewidth}p{0.55\linewidth}}
\multicolumn{1}{c}{\textbf{Research Question}} &
  \multicolumn{1}{c}{\textbf{Survey}} &
  \multicolumn{1}{c}{\textbf{Outcome}} \\ \hline
  RQ1: \rqone &
  \multicolumn{1}{c}{LES} &
  Students become more expertlike in their epistemologies associated with the role of experimentation in learning, and in research. NOMR students shift away from viewing experimental physics exclusively as a theory-testing endeavor, to one that includes a variety of important aspects of the role of experimentation in generating new knowledge. \\ \hline
  RQ2: \rqtwo &
  \multicolumn{1}{c}{PhIS} &
  NOMR labs help students develop belief in their professional capacity as experimentalists, while also helping them develop more expertlike habits of mind about experimental physics. \\ \hline
  RQ3: \rqthree &
  \multicolumn{1}{c}{FS} &
  Students become increasingly engaged with successive NOMR labs, even after the novelty wears off. They are most engaged when developing their own hypotheses with novel physics.\\
  \bottomrule
\end{tabular}%
\caption{The research questions, surveys, and outcomes of this study are summarized.}
\label{tab:results}
\end{table*}
\endgroup

\subsection{Epistemological hazards of fictitious physics}

\hl{The use of fictitious laws of physics raises concerns about whether interacting with fictitious laws of physics can negatively affect students' physical intuition and conceptual understanding. These concerns have been on our minds since the first trial of NOMR in early 2020. For that reason, we take care to maintain conceptual separation between NOMR-unique physics and the physics of our universe:}

\begin{itemize}
    \item The introduction of every NOMR lab manual (except VR1: Charge) makes clear that the physics students will be investigating was created for the purpose of the lab and does not exist in our universe. The lab manuals frame activities with fictitious physics as being original investigations into unknown physics, putting students in the shoes of Coulomb and peers.
    \item Fictitious particles are given silly names (e.g. minty particles); where they are not named in the lab manual, students are encouraged to come up with their own names for the particles. They discovered them, after all.
    \item At no point do students work with simulations of both real and fictitious physics at the same time.
    \item Fictitious particles are only ever referenced in context of the laboratory component of a course; they are not mentioned in lecture or tutorial components.
\end{itemize}

\hl{To date, we have not seen evidence of negative impacts on students' conceptual or procedural physics knowledge arising from their work with fictitious physics.}

\section{Conclusions}


Steering students away from confirmation of known facts and into a different (simulated) universe sounds like a day in the life of Ms. Frizzle's class \cite{Cole1990TheSystem}; in lieu of a magic school bus, we use VR headsets.  In both cases, students are transported to the teacher's choice of hands-on learning environment to create knowledge through collaboration with their peers in a fun, engaging, and memorable environment. In NOMR, students learn to gain knowledge in the way that experts do.


This study contributes to the physics education community's effort to create laboratory activities that foster students' growth along affective and epistemological dimensions. We have demonstrated that these goals can be achieved by inventing new physics for students to explore, effectively drawing a new frontier of physics at the introductory level.
Our findings (summarized in Table \ref{tab:results}) begin to paint a positive picture of the affective impacts of labs featuring fictitious physics, but our experience as instructors suggests that we have yet to capture the full effect of this approach. 

Subsequent analysis will include respondent-relative analysis of flow data and examine students' responses across surveys to identify correlations between change toward expertlike beliefs, engagement in lab activities, and development of self-efficacy.



That student engagement remained strong despite the decay of the novelty of VR is an important finding to consider for others interested in the role of immersive technologies in physics education.
Replicating traditional or 2D simulator-based physics instructional activities in VR has a track record of being more engaging than comparable treatments in other media, but yielding little to no educational benefit in comparison \cite{Smith2017AElectrostatics, Porter2019AMagnetostatics, Madden2018VirtualBeyond}. 
Our findings suggest that instead, educators, researchers, and developers interested in the use of immersive technology in the physics classroom should collaborate to identify niches where its affordances can be leveraged to create unique learning experiences.



Whether VR is strictly necessary to implement the fictitious-physics approach is difficult to say with scientific rigor, and while we do not attempt to do so here, we predict that it is necessary to a greater or lesser degree. That this study found positive results using a VR intervention suggests that the utility of VR in physics education lies in niches where it lets us create learning experiences that would be infeasible or impossible by other means.
In other words, an effective educational VR activity is not effective because it uses VR any more than a PhET \cite{Perkins2006PhET:Physics} is effective because it runs in a browser. Rather, it creates a learning experience tailored to solve a specific pedagogical problem, using the medium that provides the most appropriate foundation for doing so.

\section{Acknowledgements}

The authors thank Sabrina Cheng and Tyler Flom for their assistance coding the LES data, Kazumi Tolich for major contributions to NOMR's instructional materials and allowing us to administer all of these surveys in her course, David Aplin and Eddie Mendoza for heroically taking on all of the technical legwork to facilitate NOMR's classroom implementation, and Eugenia Etkina, Peter Shaffer, and Charlotte Zimmerman for their insightful feedback on early drafts of this article. This work would not have been possible without the many TAs who have taught NOMR labs and the supportive community of the UW Physics Education Research Group; both groups are too numerous to list here, but you know who you are. Thank you.

We gratefully acknowledge financial support from the UW Student Technology Fee committee and the National Science Foundation GRF (DGE-1256082).


\bibliography{references}

\begin{thebibliography}{44}%
\makeatletter
\providecommand \@ifxundefined [1]{%
 \@ifx{#1\undefined}
}%
\providecommand \@ifnum [1]{%
 \ifnum #1\expandafter \@firstoftwo
 \else \expandafter \@secondoftwo
 \fi
}%
\providecommand \@ifx [1]{%
 \ifx #1\expandafter \@firstoftwo
 \else \expandafter \@secondoftwo
 \fi
}%
\providecommand \natexlab [1]{#1}%
\providecommand \enquote  [1]{``#1''}%
\providecommand \bibnamefont  [1]{#1}%
\providecommand \bibfnamefont [1]{#1}%
\providecommand \citenamefont [1]{#1}%
\providecommand \href@noop [0]{\@secondoftwo}%
\providecommand \href [0]{\begingroup \@sanitize@url \@href}%
\providecommand \@href[1]{\@@startlink{#1}\@@href}%
\providecommand \@@href[1]{\endgroup#1\@@endlink}%
\providecommand \@sanitize@url [0]{\catcode `\\12\catcode `\$12\catcode `\&12\catcode `\#12\catcode `\^12\catcode `\_12\catcode `\%12\relax}%
\providecommand \@@startlink[1]{}%
\providecommand \@@endlink[0]{}%
\providecommand \url  [0]{\begingroup\@sanitize@url \@url }%
\providecommand \@url [1]{\endgroup\@href {#1}{\urlprefix }}%
\providecommand \urlprefix  [0]{URL }%
\providecommand \Eprint [0]{\href }%
\providecommand \doibase [0]{https://doi.org/}%
\providecommand \selectlanguage [0]{\@gobble}%
\providecommand \bibinfo  [0]{\@secondoftwo}%
\providecommand \bibfield  [0]{\@secondoftwo}%
\providecommand \translation [1]{[#1]}%
\providecommand \BibitemOpen [0]{}%
\providecommand \bibitemStop [0]{}%
\providecommand \bibitemNoStop [0]{.\EOS\space}%
\providecommand \EOS [0]{\spacefactor3000\relax}%
\providecommand \BibitemShut  [1]{\csname bibitem#1\endcsname}%
\let\auto@bib@innerbib\@empty
\bibitem [{\citenamefont {Kozminski}\ \emph {et~al.}(2014)\citenamefont {Kozminski}, \citenamefont {Lewandowski}, \citenamefont {Beverly}, \citenamefont {Lindaas}, \citenamefont {Deardorff}, \citenamefont {Reagan}, \citenamefont {Dietz}, \citenamefont {Tagg}, \citenamefont {Eblen-­-Zayas}, \citenamefont {Williams}, \citenamefont {Hobbs},\ and\ \citenamefont {Zwickl}}]{AAPTLabGuidelines}%
  \BibitemOpen
  \bibfield  {author} {\bibinfo {author} {\bibfnamefont {J.}~\bibnamefont {Kozminski}}, \bibinfo {author} {\bibfnamefont {H.}~\bibnamefont {Lewandowski}}, \bibinfo {author} {\bibfnamefont {N.}~\bibnamefont {Beverly}}, \bibinfo {author} {\bibfnamefont {S.}~\bibnamefont {Lindaas}}, \bibinfo {author} {\bibfnamefont {D.}~\bibnamefont {Deardorff}}, \bibinfo {author} {\bibfnamefont {A.}~\bibnamefont {Reagan}}, \bibinfo {author} {\bibfnamefont {R.}~\bibnamefont {Dietz}}, \bibinfo {author} {\bibfnamefont {R.}~\bibnamefont {Tagg}}, \bibinfo {author} {\bibfnamefont {M.}~\bibnamefont {Eblen-­-Zayas}}, \bibinfo {author} {\bibfnamefont {J.}~\bibnamefont {Williams}}, \bibinfo {author} {\bibfnamefont {R.}~\bibnamefont {Hobbs}},\ and\ \bibinfo {author} {\bibfnamefont {B.}~\bibnamefont {Zwickl}},\ }\href {https://www.aapt.org/Resources/upload/LabGuidlinesDocument_EBendorsed_nov10.pdf} {\emph {\bibinfo {title} {{AAPT Recommendations for the Undergraduate Physics Laboratory Curriculum}}}},\ \bibinfo {type} {Tech. Rep.}\
  (\bibinfo  {institution} {AAPT},\ \bibinfo {year} {2014})\BibitemShut {NoStop}%
\bibitem [{\citenamefont {Smith}\ and\ \citenamefont {Holmes}(2021)}]{Smith2021BestLabs}%
  \BibitemOpen
  \bibfield  {author} {\bibinfo {author} {\bibfnamefont {E.~M.}\ \bibnamefont {Smith}}\ and\ \bibinfo {author} {\bibfnamefont {N.~G.}\ \bibnamefont {Holmes}},\ }\bibfield  {title} {\bibinfo {title} {{Best practice for instructional labs}},\ }\href {https://doi.org/10.1038/s41567-021-01256-6} {\bibfield  {journal} {\bibinfo  {journal} {Nature Physics}\ ,\ \bibinfo {pages} {1}} (\bibinfo {year} {2021})}\BibitemShut {NoStop}%
\bibitem [{\citenamefont {Holmes}\ \emph {et~al.}(2017)\citenamefont {Holmes}, \citenamefont {Olsen}, \citenamefont {Thomas},\ and\ \citenamefont {Wieman}}]{Holmes2017ValueContent}%
  \BibitemOpen
  \bibfield  {author} {\bibinfo {author} {\bibfnamefont {N.~G.}\ \bibnamefont {Holmes}}, \bibinfo {author} {\bibfnamefont {J.}~\bibnamefont {Olsen}}, \bibinfo {author} {\bibfnamefont {J.~L.}\ \bibnamefont {Thomas}},\ and\ \bibinfo {author} {\bibfnamefont {C.~E.}\ \bibnamefont {Wieman}},\ }\bibfield  {title} {\bibinfo {title} {{Value added or misattributed? A multi-institution study on the educational benefit of labs for reinforcing physics content}},\ }\bibfield  {journal} {\bibinfo  {journal} {Physical Review Physics Education Research}\ }\textbf {\bibinfo {volume} {13}},\ \href {https://doi.org/10.1103/PhysRevPhysEducRes.13.010129} {10.1103/PhysRevPhysEducRes.13.010129} (\bibinfo {year} {2017})\BibitemShut {NoStop}%
\bibitem [{\citenamefont {Hu}\ \emph {et~al.}(2017)\citenamefont {Hu}, \citenamefont {Zwickl}, \citenamefont {Wilcox},\ and\ \citenamefont {Lewandowski}}]{HuPRPER2017}%
  \BibitemOpen
  \bibfield  {author} {\bibinfo {author} {\bibfnamefont {D.}~\bibnamefont {Hu}}, \bibinfo {author} {\bibfnamefont {B.~M.}\ \bibnamefont {Zwickl}}, \bibinfo {author} {\bibfnamefont {B.~R.}\ \bibnamefont {Wilcox}},\ and\ \bibinfo {author} {\bibfnamefont {H.~J.}\ \bibnamefont {Lewandowski}},\ }\bibfield  {title} {\bibinfo {title} {{Qualitative investigation of students' views about experimental physics}},\ }\bibfield  {journal} {\bibinfo  {journal} {Physical Review Physics Education Research}\ }\textbf {\bibinfo {volume} {13}},\ \href {https://doi.org/10.1103/PhysRevPhysEducRes.13.020134} {10.1103/PhysRevPhysEducRes.13.020134} (\bibinfo {year} {2017})\BibitemShut {NoStop}%
\bibitem [{\citenamefont {Etkina}\ and\ \citenamefont {Heuvelen}(2007)}]{Etkina2007InvestigativePhysics}%
  \BibitemOpen
  \bibfield  {author} {\bibinfo {author} {\bibfnamefont {E.}~\bibnamefont {Etkina}}\ and\ \bibinfo {author} {\bibfnamefont {A.~V.}\ \bibnamefont {Heuvelen}},\ }\bibfield  {title} {\bibinfo {title} {{Investigative Science Learning Environment - A Science Process Approach to Learning Physics}},\ }in\ \href@noop {} {\emph {\bibinfo {booktitle} {Research-Based Reform of University Physics}}},\ Vol.~\bibinfo {volume} {1}\ (\bibinfo {year} {2007})\BibitemShut {NoStop}%
\bibitem [{\citenamefont {Bandura}(1977)}]{Bandura1977Self-efficacy:Change.}%
  \BibitemOpen
  \bibfield  {author} {\bibinfo {author} {\bibfnamefont {A.}~\bibnamefont {Bandura}},\ }\bibfield  {title} {\bibinfo {title} {{Self-efficacy: Toward a unifying theory of behavioral change.}},\ }\href {https://doi.org/10.1037/0033-295X.84.2.191} {\bibfield  {journal} {\bibinfo  {journal} {Psychological Review}\ }\textbf {\bibinfo {volume} {84}},\ \bibinfo {pages} {191} (\bibinfo {year} {1977})}\BibitemShut {NoStop}%
\bibitem [{\citenamefont {Etkina}\ \emph {et~al.}(2006{\natexlab{a}})\citenamefont {Etkina}, \citenamefont {Van~Heuvelen}, \citenamefont {White-Brahmia}, \citenamefont {Brookes}, \citenamefont {Gentile}, \citenamefont {Murthy}, \citenamefont {Rosengrant},\ and\ \citenamefont {Warren}}]{Etkina2006ScientificAssessment}%
  \BibitemOpen
  \bibfield  {author} {\bibinfo {author} {\bibfnamefont {E.}~\bibnamefont {Etkina}}, \bibinfo {author} {\bibfnamefont {A.}~\bibnamefont {Van~Heuvelen}}, \bibinfo {author} {\bibfnamefont {S.}~\bibnamefont {White-Brahmia}}, \bibinfo {author} {\bibfnamefont {D.~T.}\ \bibnamefont {Brookes}}, \bibinfo {author} {\bibfnamefont {M.}~\bibnamefont {Gentile}}, \bibinfo {author} {\bibfnamefont {S.}~\bibnamefont {Murthy}}, \bibinfo {author} {\bibfnamefont {D.}~\bibnamefont {Rosengrant}},\ and\ \bibinfo {author} {\bibfnamefont {A.}~\bibnamefont {Warren}},\ }\bibfield  {title} {\bibinfo {title} {{Scientific abilities and their assessment}},\ }\bibfield  {journal} {\bibinfo  {journal} {Physical Review Special Topics - Physics Education Research}\ }\textbf {\bibinfo {volume} {2}},\ \href {https://doi.org/10.1103/PhysRevSTPER.2.020103} {10.1103/PhysRevSTPER.2.020103} (\bibinfo {year} {2006}{\natexlab{a}})\BibitemShut {NoStop}%
\bibitem [{\citenamefont {Freeman}\ \emph {et~al.}(2014)\citenamefont {Freeman}, \citenamefont {Eddy}, \citenamefont {McDonough}, \citenamefont {Smith}, \citenamefont {Okoroafor}, \citenamefont {Jordt},\ and\ \citenamefont {Wenderoth}}]{Freeman2014ActiveMathematics}%
  \BibitemOpen
  \bibfield  {author} {\bibinfo {author} {\bibfnamefont {S.}~\bibnamefont {Freeman}}, \bibinfo {author} {\bibfnamefont {S.~L.}\ \bibnamefont {Eddy}}, \bibinfo {author} {\bibfnamefont {M.}~\bibnamefont {McDonough}}, \bibinfo {author} {\bibfnamefont {M.~K.}\ \bibnamefont {Smith}}, \bibinfo {author} {\bibfnamefont {N.}~\bibnamefont {Okoroafor}}, \bibinfo {author} {\bibfnamefont {H.}~\bibnamefont {Jordt}},\ and\ \bibinfo {author} {\bibfnamefont {M.~P.}\ \bibnamefont {Wenderoth}},\ }\bibfield  {title} {\bibinfo {title} {{Active learning increases student performance in science, engineering, and mathematics}},\ }\href {https://doi.org/10.1073/pnas.1319030111} {\bibfield  {journal} {\bibinfo  {journal} {Proceedings of the National Academy of Sciences}\ }\textbf {\bibinfo {volume} {111}},\ \bibinfo {pages} {8410} (\bibinfo {year} {2014})}\BibitemShut {NoStop}%
\bibitem [{\citenamefont {Trumper}(2003)}]{Trumper2003ThePerspectives}%
  \BibitemOpen
  \bibfield  {author} {\bibinfo {author} {\bibfnamefont {R.}~\bibnamefont {Trumper}},\ }\bibfield  {title} {\bibinfo {title} {{The Physics Laboratory - A Historical Overview and Future Perspectives}},\ }\href {https://doi.org/10.1023/A:1025692409001} {\bibfield  {journal} {\bibinfo  {journal} {Science and Education}\ }\textbf {\bibinfo {volume} {12}},\ \bibinfo {pages} {645} (\bibinfo {year} {2003})}\BibitemShut {NoStop}%
\bibitem [{\citenamefont {Brookes}\ \emph {et~al.}(2020{\natexlab{a}})\citenamefont {Brookes}, \citenamefont {Etkina},\ and\ \citenamefont {Planinsic}}]{Brookes2020ImplementingLearning}%
  \BibitemOpen
  \bibfield  {author} {\bibinfo {author} {\bibfnamefont {D.~T.}\ \bibnamefont {Brookes}}, \bibinfo {author} {\bibfnamefont {E.}~\bibnamefont {Etkina}},\ and\ \bibinfo {author} {\bibfnamefont {G.}~\bibnamefont {Planinsic}},\ }\bibfield  {title} {\bibinfo {title} {{Implementing an epistemologically authentic approach to student-centered inquiry learning}},\ }\href {https://doi.org/10.1103/PhysRevPhysEducRes.16.020148} {\bibfield  {journal} {\bibinfo  {journal} {Physical Review Physics Education Research}\ }\textbf {\bibinfo {volume} {16}},\ \bibinfo {pages} {20148} (\bibinfo {year} {2020}{\natexlab{a}})}\BibitemShut {NoStop}%
\bibitem [{\citenamefont {Etkina}\ \emph {et~al.}(2021)\citenamefont {Etkina}, \citenamefont {Brookes},\ and\ \citenamefont {Planinsic}}]{Etkina2021ThePhysics}%
  \BibitemOpen
  \bibfield  {author} {\bibinfo {author} {\bibfnamefont {E.}~\bibnamefont {Etkina}}, \bibinfo {author} {\bibfnamefont {D.~T.}\ \bibnamefont {Brookes}},\ and\ \bibinfo {author} {\bibfnamefont {G.}~\bibnamefont {Planinsic}},\ }\bibfield  {title} {\bibinfo {title} {{The Investigative Science Learning Environment (ISLE) approach to learning physics}},\ }in\ \href {https://doi.org/10.1088/1742-6596/1882/1/012001} {\emph {\bibinfo {booktitle} {Journal of Physics: Conference Series}}},\ Vol.\ \bibinfo {volume} {1882}\ (\bibinfo  {publisher} {IOP Publishing Ltd},\ \bibinfo {year} {2021})\BibitemShut {NoStop}%
\bibitem [{\citenamefont {Etkina}(2015)}]{Etkina2015MillikanPractices}%
  \BibitemOpen
  \bibfield  {author} {\bibinfo {author} {\bibfnamefont {E.}~\bibnamefont {Etkina}},\ }\bibfield  {title} {\bibinfo {title} {{Millikan award lecture: Students of physics - Listeners, observers, or collaborative participants in physics scientific practices?}},\ }\href {https://doi.org/10.1119/1.4923432} {\bibfield  {journal} {\bibinfo  {journal} {American Journal of Physics}\ }\textbf {\bibinfo {volume} {83}},\ \bibinfo {pages} {669} (\bibinfo {year} {2015})}\BibitemShut {NoStop}%
\bibitem [{\citenamefont {Etkina}\ and\ \citenamefont {Planinsic}(2015)}]{Etkina2015DefiningPhenomenon}%
  \BibitemOpen
  \bibfield  {author} {\bibinfo {author} {\bibfnamefont {E.}~\bibnamefont {Etkina}}\ and\ \bibinfo {author} {\bibfnamefont {G.}~\bibnamefont {Planinsic}},\ }\bibfield  {title} {\bibinfo {title} {{Defining and Developing 'Critical Thinking' Through Devising and Testing Multiple Explanations of the Same Phenomenon}},\ }\href {https://doi.org/10.1119/1.4931014} {\bibfield  {journal} {\bibinfo  {journal} {The Physics Teacher}\ }\textbf {\bibinfo {volume} {53}},\ \bibinfo {pages} {432} (\bibinfo {year} {2015})}\BibitemShut {NoStop}%
\bibitem [{\citenamefont {Brookes}\ \emph {et~al.}(2020{\natexlab{b}})\citenamefont {Brookes}, \citenamefont {Etkina},\ and\ \citenamefont {Planinsic}}]{isle2020}%
  \BibitemOpen
  \bibfield  {author} {\bibinfo {author} {\bibfnamefont {D.~T.}\ \bibnamefont {Brookes}}, \bibinfo {author} {\bibfnamefont {E.}~\bibnamefont {Etkina}},\ and\ \bibinfo {author} {\bibfnamefont {G.}~\bibnamefont {Planinsic}},\ }\bibfield  {title} {\bibinfo {title} {{Implementing an epistemologically authentic approach to student-centered inquiry learning}},\ }\bibfield  {journal} {\bibinfo  {journal} {Physical Review Physics Education Research}\ }\textbf {\bibinfo {volume} {16}},\ \href {https://doi.org/10.1103/physrevphyseducres.16.020148} {10.1103/physrevphyseducres.16.020148} (\bibinfo {year} {2020}{\natexlab{b}})\BibitemShut {NoStop}%
\bibitem [{\citenamefont {Etkina}\ \emph {et~al.}(2006{\natexlab{b}})\citenamefont {Etkina}, \citenamefont {Warren},\ and\ \citenamefont {Gentile}}]{Etkina2006TheInstruction}%
  \BibitemOpen
  \bibfield  {author} {\bibinfo {author} {\bibfnamefont {E.}~\bibnamefont {Etkina}}, \bibinfo {author} {\bibfnamefont {A.}~\bibnamefont {Warren}},\ and\ \bibinfo {author} {\bibfnamefont {M.}~\bibnamefont {Gentile}},\ }\bibfield  {title} {\bibinfo {title} {{The Role of Models in Physics Instruction}},\ }\href {https://doi.org/10.1119/1.2150757} {\bibfield  {journal} {\bibinfo  {journal} {The Physics Teacher}\ }\textbf {\bibinfo {volume} {44}},\ \bibinfo {pages} {34} (\bibinfo {year} {2006}{\natexlab{b}})}\BibitemShut {NoStop}%
\bibitem [{\citenamefont {Stein}\ \emph {et~al.}(2018)\citenamefont {Stein}, \citenamefont {Smith},\ and\ \citenamefont {Holmes}}]{Stein2018ConfirmingLabs}%
  \BibitemOpen
  \bibfield  {author} {\bibinfo {author} {\bibfnamefont {M.~M.}\ \bibnamefont {Stein}}, \bibinfo {author} {\bibfnamefont {E.~M.}\ \bibnamefont {Smith}},\ and\ \bibinfo {author} {\bibfnamefont {N.~G.}\ \bibnamefont {Holmes}},\ }\bibfield  {title} {\bibinfo {title} {{Confirming what we know: Understanding questionable research practices in intro physics labs}},\ }in\ \href {https://doi.org/10.1119/perc.2018.pr.Stein} {\emph {\bibinfo {booktitle} {PERC Proceedings}}}\ (\bibinfo {address} {Washington, DC},\ \bibinfo {year} {2018})\BibitemShut {NoStop}%
\bibitem [{\citenamefont {Canright}\ \emph {et~al.}(2020)\citenamefont {Canright}, \citenamefont {Olsen},\ and\ \citenamefont {Brahmia}}]{Canright2020}%
  \BibitemOpen
  \bibfield  {author} {\bibinfo {author} {\bibfnamefont {J.~P.}\ \bibnamefont {Canright}}, \bibinfo {author} {\bibfnamefont {J.~R.}\ \bibnamefont {Olsen}},\ and\ \bibinfo {author} {\bibfnamefont {S.~W.}\ \bibnamefont {Brahmia}},\ }\bibfield  {title} {\bibinfo {title} {{Leveraging virtual reality for student development of force models in the introductory lab}},\ }in\ \href {https://doi.org/10.1119/perc.2020.pr.canright} {\emph {\bibinfo {booktitle} {Physics Education Research Conference}}}\ (\bibinfo  {publisher} {American Association of Physics Teachers (AAPT)},\ \bibinfo {year} {2020})\ pp.\ \bibinfo {pages} {75--80}\BibitemShut {NoStop}%
\bibitem [{\citenamefont {Canright}\ and\ \citenamefont {White~Brahmia}(2021)}]{Canright2021DevelopingReality}%
  \BibitemOpen
  \bibfield  {author} {\bibinfo {author} {\bibfnamefont {J.~P.}\ \bibnamefont {Canright}}\ and\ \bibinfo {author} {\bibfnamefont {S.}~\bibnamefont {White~Brahmia}},\ }\bibfield  {title} {\bibinfo {title} {{Developing expertlike epistemologies about physics empirical discovery using virtual reality}}\ }(\bibinfo  {publisher} {American Association of Physics Teachers (AAPT)},\ \bibinfo {year} {2021})\ pp.\ \bibinfo {pages} {75--80}\BibitemShut {NoStop}%
\bibitem [{\citenamefont {Walsh}\ \emph {et~al.}(2019)\citenamefont {Walsh}, \citenamefont {Quinn}, \citenamefont {Wieman},\ and\ \citenamefont {Holmes}}]{Walsh2019QuantifyingThinking}%
  \BibitemOpen
  \bibfield  {author} {\bibinfo {author} {\bibfnamefont {C.}~\bibnamefont {Walsh}}, \bibinfo {author} {\bibfnamefont {K.~N.}\ \bibnamefont {Quinn}}, \bibinfo {author} {\bibfnamefont {C.}~\bibnamefont {Wieman}},\ and\ \bibinfo {author} {\bibfnamefont {N.~G.}\ \bibnamefont {Holmes}},\ }\bibfield  {title} {\bibinfo {title} {{Quantifying critical thinking: Development and validation of the physics lab inventory of critical thinking}},\ }\bibfield  {journal} {\bibinfo  {journal} {Physical Review Physics Education Research}\ }\textbf {\bibinfo {volume} {15}},\ \href {https://doi.org/10.1103/physrevphyseducres.15.010135} {10.1103/physrevphyseducres.15.010135} (\bibinfo {year} {2019})\BibitemShut {NoStop}%
\bibitem [{\citenamefont {Vignal}\ \emph {et~al.}(2023)\citenamefont {Vignal}, \citenamefont {Geschwind}, \citenamefont {Pollard}, \citenamefont {Henderson}, \citenamefont {Caballero},\ and\ \citenamefont {Lewandowski}}]{Vignal2023SurveyLabs}%
  \BibitemOpen
  \bibfield  {author} {\bibinfo {author} {\bibfnamefont {M.}~\bibnamefont {Vignal}}, \bibinfo {author} {\bibfnamefont {G.}~\bibnamefont {Geschwind}}, \bibinfo {author} {\bibfnamefont {B.}~\bibnamefont {Pollard}}, \bibinfo {author} {\bibfnamefont {R.}~\bibnamefont {Henderson}}, \bibinfo {author} {\bibfnamefont {M.~D.}\ \bibnamefont {Caballero}},\ and\ \bibinfo {author} {\bibfnamefont {H.~J.}\ \bibnamefont {Lewandowski}},\ }\href@noop {} {\bibinfo {title} {{Survey of physics reasoning on uncertainty concepts in experiments: an assessment of measurement uncertainty for introductory physics labs}}} (\bibinfo {year} {2023})\BibitemShut {NoStop}%
\bibitem [{\citenamefont {Hu}\ and\ \citenamefont {Zwickl}(2018{\natexlab{a}})}]{HuZwickl2017}%
  \BibitemOpen
  \bibfield  {author} {\bibinfo {author} {\bibfnamefont {D.}~\bibnamefont {Hu}}\ and\ \bibinfo {author} {\bibfnamefont {B.~M.}\ \bibnamefont {Zwickl}},\ }\bibfield  {title} {\bibinfo {title} {{Examining students' personal epistemology: the role of physics experiments and relation with theory}},\ }in\ \href {https://doi.org/10.1119/perc.2017.juried.003} {\emph {\bibinfo {booktitle} {PERC Proceedings}}}\ (\bibinfo  {publisher} {American Association of Physics Teachers (AAPT)},\ \bibinfo {year} {2018})\ pp.\ \bibinfo {pages} {11--14}\BibitemShut {NoStop}%
\bibitem [{\citenamefont {Hu}\ and\ \citenamefont {Zwickl}(2018{\natexlab{b}})}]{Hu2018ExaminingStudents}%
  \BibitemOpen
  \bibfield  {author} {\bibinfo {author} {\bibfnamefont {D.}~\bibnamefont {Hu}}\ and\ \bibinfo {author} {\bibfnamefont {B.~M.}\ \bibnamefont {Zwickl}},\ }\bibfield  {title} {\bibinfo {title} {{Examining students' views about validity of experiments: From introductory to Ph.D. students}},\ }\bibfield  {journal} {\bibinfo  {journal} {Physical Review Physics Education Research}\ }\textbf {\bibinfo {volume} {14}},\ \href {https://doi.org/10.1103/PhysRevPhysEducRes.14.010121} {10.1103/PhysRevPhysEducRes.14.010121} (\bibinfo {year} {2018}{\natexlab{b}})\BibitemShut {NoStop}%
\bibitem [{\citenamefont {Reilly}\ \emph {et~al.}(2021)\citenamefont {Reilly}, \citenamefont {McGivney}, \citenamefont {Dede},\ and\ \citenamefont {Grotzer}}]{Reilly2021AssessingApproach}%
  \BibitemOpen
  \bibfield  {author} {\bibinfo {author} {\bibfnamefont {J.~M.}\ \bibnamefont {Reilly}}, \bibinfo {author} {\bibfnamefont {E.}~\bibnamefont {McGivney}}, \bibinfo {author} {\bibfnamefont {C.}~\bibnamefont {Dede}},\ and\ \bibinfo {author} {\bibfnamefont {T.}~\bibnamefont {Grotzer}},\ }\bibfield  {title} {\bibinfo {title} {{Assessing Science Identity Exploration in Immersive Virtual Environments: A Mixed Methods Approach}},\ }\href {https://doi.org/10.1080/00220973.2020.1712313} {\bibfield  {journal} {\bibinfo  {journal} {Journal of Experimental Education}\ }\textbf {\bibinfo {volume} {89}},\ \bibinfo {pages} {468} (\bibinfo {year} {2021})}\BibitemShut {NoStop}%
\bibitem [{\citenamefont {Hazari}\ \emph {et~al.}()\citenamefont {Hazari}, \citenamefont {Sadler},\ and\ \citenamefont {Sonnert}}]{HazariTheEthnicity}%
  \BibitemOpen
  \bibfield  {author} {\bibinfo {author} {\bibfnamefont {Z.}~\bibnamefont {Hazari}}, \bibinfo {author} {\bibfnamefont {P.~M.}\ \bibnamefont {Sadler}},\ and\ \bibinfo {author} {\bibfnamefont {G.}~\bibnamefont {Sonnert}},\ }\bibfield  {title} {\bibinfo {title} {{The Science Identity of College Students: Exploring the Intersection of Gender, Race, and Ethnicity}},\ }\href {http://www.jstor.org/stable/43631586} {\bibfield  {journal} {\bibinfo  {journal} {Source: Journal of College Science Teaching}\ }\textbf {\bibinfo {volume} {42}},\ \bibinfo {pages} {82}}\BibitemShut {NoStop}%
\bibitem [{\citenamefont {Adams}\ and\ \citenamefont {Wieman}(2011)}]{Adams2011DevelopmentThinking}%
  \BibitemOpen
  \bibfield  {author} {\bibinfo {author} {\bibfnamefont {W.~K.}\ \bibnamefont {Adams}}\ and\ \bibinfo {author} {\bibfnamefont {C.~E.}\ \bibnamefont {Wieman}},\ }\bibfield  {title} {\bibinfo {title} {{Development and Validation of Instruments to Measure Learning of Expert-Like Thinking}},\ }\href {https://doi.org/10.1080/09500693.2010.512369} {\bibfield  {journal} {\bibinfo  {journal} {International Journal of Science Education}\ }\textbf {\bibinfo {volume} {33}},\ \bibinfo {pages} {1289} (\bibinfo {year} {2011})}\BibitemShut {NoStop}%
\bibitem [{\citenamefont {Csikszentmihalyi}(1990)}]{Csikszentmihalyi1990Flow:Experience}%
  \BibitemOpen
  \bibfield  {author} {\bibinfo {author} {\bibfnamefont {M.}~\bibnamefont {Csikszentmihalyi}},\ }\bibfield  {title} {\bibinfo {title} {{Flow: The Psychology of Optimal Experience}}\ }(\bibinfo {year} {1990})\BibitemShut {NoStop}%
\bibitem [{\citenamefont {Csikszentmihalyi}(2014)}]{Csikszentmihalyi2014ApplicationsCsikszentmihalyi.}%
  \BibitemOpen
  \bibfield  {author} {\bibinfo {author} {\bibfnamefont {M.}~\bibnamefont {Csikszentmihalyi}},\ }\href {https://doi.org/10.1007/978-94-017-9094-9} {\emph {\bibinfo {title} {Applications of flow in human development and education: The collected works of Mihaly Csikszentmihalyi.}}}\ (\bibinfo  {publisher} {Springer Science + Business Media},\ \bibinfo {address} {New York, NY, US},\ \bibinfo {year} {2014})\ pp.\ \bibinfo {pages} {494, xxii, 494--xxii}\BibitemShut {NoStop}%
\bibitem [{\citenamefont {Rebello}\ and\ \citenamefont {Zollman}(2013)}]{Rebello2013ProblemFlow}%
  \BibitemOpen
  \bibfield  {author} {\bibinfo {author} {\bibfnamefont {N.~S.}\ \bibnamefont {Rebello}}\ and\ \bibinfo {author} {\bibfnamefont {D.}~\bibnamefont {Zollman}},\ }\bibfield  {title} {\bibinfo {title} {{Problem Solving and Motivation - Getting our Students in Flow}},\ }in\ \href@noop {} {\emph {\bibinfo {booktitle} {Physics Education Research Conference 2013}}},\ \bibinfo {series and number} {PER Conference Invited Paper}\ (\bibinfo {address} {Portland, OR},\ \bibinfo {year} {2013})\ pp.\ \bibinfo {pages} {39--41}\BibitemShut {NoStop}%
\bibitem [{\citenamefont {Karelina}\ \emph {et~al.}(2022)\citenamefont {Karelina}, \citenamefont {Etkina}, \citenamefont {Bohacek}, \citenamefont {Vonk}, \citenamefont {Kagan}, \citenamefont {Warren},\ and\ \citenamefont {Brookes}}]{Karelina2022ComparingActivities}%
  \BibitemOpen
  \bibfield  {author} {\bibinfo {author} {\bibfnamefont {A.}~\bibnamefont {Karelina}}, \bibinfo {author} {\bibfnamefont {E.}~\bibnamefont {Etkina}}, \bibinfo {author} {\bibfnamefont {P.}~\bibnamefont {Bohacek}}, \bibinfo {author} {\bibfnamefont {M.}~\bibnamefont {Vonk}}, \bibinfo {author} {\bibfnamefont {M.}~\bibnamefont {Kagan}}, \bibinfo {author} {\bibfnamefont {A.~R.}\ \bibnamefont {Warren}},\ and\ \bibinfo {author} {\bibfnamefont {D.~T.}\ \bibnamefont {Brookes}},\ }\bibfield  {title} {\bibinfo {title} {{Comparing students' flow states during apparatus-based versus video-based lab activities}},\ }\bibfield  {journal} {\bibinfo  {journal} {European Journal of Physics}\ }\textbf {\bibinfo {volume} {43}},\ \href {https://doi.org/10.1088/1361-6404/ac683f} {10.1088/1361-6404/ac683f} (\bibinfo {year} {2022})\BibitemShut {NoStop}%
\bibitem [{\citenamefont {Zaretskii}(2009)}]{Zaretskii2009TheDevelopment}%
  \BibitemOpen
  \bibfield  {author} {\bibinfo {author} {\bibfnamefont {V.~K.}\ \bibnamefont {Zaretskii}},\ }\bibfield  {title} {\bibinfo {title} {{The Zone of Proximal Development}},\ }\href {https://doi.org/10.2753/RPO1061-0405470604} {\bibfield  {journal} {\bibinfo  {journal} {Journal of Russian {\&} East European Psychology}\ }\textbf {\bibinfo {volume} {47}},\ \bibinfo {pages} {70} (\bibinfo {year} {2009})}\BibitemShut {NoStop}%
\bibitem [{\citenamefont {Schwartz}\ \emph {et~al.}(2005)\citenamefont {Schwartz}, \citenamefont {Bransford}, \citenamefont {Sears},\ and\ \citenamefont {{others}}}]{Schwartz2005EfficiencyTransfer}%
  \BibitemOpen
  \bibfield  {author} {\bibinfo {author} {\bibfnamefont {D.~L.}\ \bibnamefont {Schwartz}}, \bibinfo {author} {\bibfnamefont {J.~D.}\ \bibnamefont {Bransford}}, \bibinfo {author} {\bibfnamefont {D.}~\bibnamefont {Sears}},\ and\ \bibinfo {author} {\bibnamefont {{others}}},\ }\bibfield  {title} {\bibinfo {title} {{Efficiency and innovation in transfer}},\ }\href@noop {} {\bibfield  {journal} {\bibinfo  {journal} {Transfer of learning from a modern multidisciplinary perspective}\ }\textbf {\bibinfo {volume} {3}},\ \bibinfo {pages} {1} (\bibinfo {year} {2005})}\BibitemShut {NoStop}%
\bibitem [{\citenamefont {Massimini}\ \emph {et~al.}(1987)\citenamefont {Massimini}, \citenamefont {Csikszentmihalyi},\ and\ \citenamefont {Carli}}]{Massimini1987TheRehabilitation}%
  \BibitemOpen
  \bibfield  {author} {\bibinfo {author} {\bibfnamefont {F.}~\bibnamefont {Massimini}}, \bibinfo {author} {\bibfnamefont {M.}~\bibnamefont {Csikszentmihalyi}},\ and\ \bibinfo {author} {\bibfnamefont {M.}~\bibnamefont {Carli}},\ }\bibfield  {title} {\bibinfo {title} {{The Monitoring of Optimal Experience A Tool for Psychiatric Rehabilitation}},\ }\href {https://journals.lww.com/jonmd/Fulltext/1987/09000/The_Monitoring_of_Optimal_Experience_A_Tool_for.6.aspx} {\bibfield  {journal} {\bibinfo  {journal} {The Journal of Nervous and Mental Disease}\ }\textbf {\bibinfo {volume} {175}} (\bibinfo {year} {1987})}\BibitemShut {NoStop}%
\bibitem [{\citenamefont {Massimini}\ and\ \citenamefont {Carli}(1988)}]{Massimini1988OptimalConsciousness.}%
  \BibitemOpen
  \bibfield  {author} {\bibinfo {author} {\bibfnamefont {F.}~\bibnamefont {Massimini}}\ and\ \bibinfo {author} {\bibfnamefont {M.}~\bibnamefont {Carli}},\ }\href@noop {} {\emph {\bibinfo {title} {Optimal experience: Psychological studies of flow in consciousness.}}},\ edited by\ \bibinfo {editor} {\bibfnamefont {M.}~\bibnamefont {Csikszentmihalyi}}\ and\ \bibinfo {editor} {\bibfnamefont {I.~S.}\ \bibnamefont {Csikszentmihalyi}}\ (\bibinfo  {publisher} {Cambridge University Press},\ \bibinfo {address} {New York, NY, US},\ \bibinfo {year} {1988})\ pp.\ \bibinfo {pages} {266--287}\BibitemShut {NoStop}%
\bibitem [{\citenamefont {Jackson}\ and\ \citenamefont {Marsh}(1996)}]{Jackson1996DevelopmentScale}%
  \BibitemOpen
  \bibfield  {author} {\bibinfo {author} {\bibfnamefont {S.~A.}\ \bibnamefont {Jackson}}\ and\ \bibinfo {author} {\bibfnamefont {H.~W.}\ \bibnamefont {Marsh}},\ }\bibfield  {title} {\bibinfo {title} {{Development and Validation of a Scale to Measure Optimal Experience: The Flow State Scale}},\ }\href {https://doi.org/10.1123/jsep.18.1.17} {\bibfield  {journal} {\bibinfo  {journal} {Journal of Sport and Exercise Psychology}\ }\textbf {\bibinfo {volume} {18}},\ \bibinfo {pages} {17 } (\bibinfo {year} {1996})}\BibitemShut {NoStop}%
\bibitem [{\citenamefont {Day}\ \emph {et~al.}(2016)\citenamefont {Day}, \citenamefont {Stang}, \citenamefont {Holmes}, \citenamefont {Kumar},\ and\ \citenamefont {Bonn}}]{Day2016GenderLaboratory}%
  \BibitemOpen
  \bibfield  {author} {\bibinfo {author} {\bibfnamefont {J.}~\bibnamefont {Day}}, \bibinfo {author} {\bibfnamefont {J.~B.}\ \bibnamefont {Stang}}, \bibinfo {author} {\bibfnamefont {N.~G.}\ \bibnamefont {Holmes}}, \bibinfo {author} {\bibfnamefont {D.}~\bibnamefont {Kumar}},\ and\ \bibinfo {author} {\bibfnamefont {D.~A.}\ \bibnamefont {Bonn}},\ }\bibfield  {title} {\bibinfo {title} {{Gender gaps and gendered action in a first-year physics laboratory}},\ }\href@noop {} {\bibfield  {journal} {\bibinfo  {journal} {Physical Review Physics Education Research}\ }\textbf {\bibinfo {volume} {12}},\ \bibinfo {pages} {20104} (\bibinfo {year} {2016})}\BibitemShut {NoStop}%
\bibitem [{\citenamefont {Holmes}\ \emph {et~al.}(2022)\citenamefont {Holmes}, \citenamefont {Heath}, \citenamefont {Hubenig}, \citenamefont {Jeon}, \citenamefont {Kalender}, \citenamefont {Stump},\ and\ \citenamefont {Sayre}}]{Holmes2022EvaluatingEquity}%
  \BibitemOpen
  \bibfield  {author} {\bibinfo {author} {\bibfnamefont {N.}~\bibnamefont {Holmes}}, \bibinfo {author} {\bibfnamefont {G.}~\bibnamefont {Heath}}, \bibinfo {author} {\bibfnamefont {K.}~\bibnamefont {Hubenig}}, \bibinfo {author} {\bibfnamefont {S.}~\bibnamefont {Jeon}}, \bibinfo {author} {\bibfnamefont {Z.~Y.}\ \bibnamefont {Kalender}}, \bibinfo {author} {\bibfnamefont {E.}~\bibnamefont {Stump}},\ and\ \bibinfo {author} {\bibfnamefont {E.~C.}\ \bibnamefont {Sayre}},\ }\bibfield  {title} {\bibinfo {title} {{Evaluating the role of student preference in physics lab group equity}},\ }\href {https://doi.org/10.1103/PhysRevPhysEducRes.18.010106} {\bibfield  {journal} {\bibinfo  {journal} {Physical Review Physics Education Research}\ }\textbf {\bibinfo {volume} {18}},\ \bibinfo {pages} {10106} (\bibinfo {year} {2022})}\BibitemShut {NoStop}%
\bibitem [{\citenamefont {Wilcoxon}(1945)}]{Wilcoxon1945IndividualMethods}%
  \BibitemOpen
  \bibfield  {author} {\bibinfo {author} {\bibfnamefont {F.}~\bibnamefont {Wilcoxon}},\ }\bibfield  {title} {\bibinfo {title} {{Individual Comparisons by Ranking Methods}},\ }\href {https://doi.org/10.2307/3001968} {\bibfield  {journal} {\bibinfo  {journal} {Biometrics Bulletin}\ }\textbf {\bibinfo {volume} {1}},\ \bibinfo {pages} {80} (\bibinfo {year} {1945})}\BibitemShut {NoStop}%
\bibitem [{\citenamefont {Clark}(1983)}]{Clark1983ReconsideringMedia}%
  \BibitemOpen
  \bibfield  {author} {\bibinfo {author} {\bibfnamefont {R.~E.}\ \bibnamefont {Clark}},\ }\bibfield  {title} {\bibinfo {title} {{Reconsidering research on learning from media}},\ }\href@noop {} {\bibfield  {journal} {\bibinfo  {journal} {Review of educational research}\ }\textbf {\bibinfo {volume} {53}},\ \bibinfo {pages} {445} (\bibinfo {year} {1983})}\BibitemShut {NoStop}%
\bibitem [{\citenamefont {McDonnell}\ and\ \citenamefont {Milton}(2014)}]{McDonnell2014GoingStates}%
  \BibitemOpen
  \bibfield  {author} {\bibinfo {author} {\bibfnamefont {A.}~\bibnamefont {McDonnell}}\ and\ \bibinfo {author} {\bibfnamefont {D.}~\bibnamefont {Milton}},\ }\bibfield  {title} {\bibinfo {title} {{Going with the flow: reconsidering 'repetitive behaviour' through the concept of 'flow states'}},\ }in\ \href {https://kar.kent.ac.uk/62647/} {\emph {\bibinfo {booktitle} {Good Autism Practice: autism, happiness and wellbeing}}},\ \bibinfo {editor} {edited by\ \bibinfo {editor} {\bibfnamefont {G.}~\bibnamefont {Jones}}\ and\ \bibinfo {editor} {\bibfnamefont {E.}~\bibnamefont {Hurley}}}\ (\bibinfo  {publisher} {BILD},\ \bibinfo {address} {Birmingham, UK},\ \bibinfo {year} {2014})\ pp.\ \bibinfo {pages} {38--47}\BibitemShut {NoStop}%
\bibitem [{\citenamefont {Cole}\ and\ \citenamefont {Degen}(1990)}]{Cole1990TheSystem}%
  \BibitemOpen
  \bibfield  {author} {\bibinfo {author} {\bibfnamefont {J.}~\bibnamefont {Cole}}\ and\ \bibinfo {author} {\bibfnamefont {B.}~\bibnamefont {Degen}},\ }\href@noop {} {\emph {\bibinfo {title} {{The Magic School Bus Lost in the Solar System}}}}\ (\bibinfo  {publisher} {Scholastic},\ \bibinfo {year} {1990})\BibitemShut {NoStop}%
\bibitem [{\citenamefont {Smith}\ \emph {et~al.}(2017)\citenamefont {Smith}, \citenamefont {Byrum}, \citenamefont {McCormick}, \citenamefont {Young}, \citenamefont {Orban},\ and\ \citenamefont {Porter}}]{Smith2017AElectrostatics}%
  \BibitemOpen
  \bibfield  {author} {\bibinfo {author} {\bibfnamefont {J.~R.}\ \bibnamefont {Smith}}, \bibinfo {author} {\bibfnamefont {A.}~\bibnamefont {Byrum}}, \bibinfo {author} {\bibfnamefont {T.~M.}\ \bibnamefont {McCormick}}, \bibinfo {author} {\bibfnamefont {N.}~\bibnamefont {Young}}, \bibinfo {author} {\bibfnamefont {C.}~\bibnamefont {Orban}},\ and\ \bibinfo {author} {\bibfnamefont {C.~D.}\ \bibnamefont {Porter}},\ }\bibfield  {title} {\bibinfo {title} {{A controlled study of stereoscopic virtual reality in freshman electrostatics}},\ }in\ \href {https://doi.org/10.1119/perc.2017.pr.089} {\emph {\bibinfo {booktitle} {PERC Proceedings}}}\ (\bibinfo {address} {Cincinnati, OH},\ \bibinfo {year} {2017})\ pp.\ \bibinfo {pages} {376--379}\BibitemShut {NoStop}%
\bibitem [{\citenamefont {Porter}\ \emph {et~al.}(2019)\citenamefont {Porter}, \citenamefont {Brown}, \citenamefont {Smith}, \citenamefont {Stagar}, \citenamefont {Simmons}, \citenamefont {Nieberding}, \citenamefont {Ayers},\ and\ \citenamefont {Orban}}]{Porter2019AMagnetostatics}%
  \BibitemOpen
  \bibfield  {author} {\bibinfo {author} {\bibfnamefont {C.~D.}\ \bibnamefont {Porter}}, \bibinfo {author} {\bibfnamefont {J.~R.}\ \bibnamefont {Brown}}, \bibinfo {author} {\bibfnamefont {J.~R.}\ \bibnamefont {Smith}}, \bibinfo {author} {\bibfnamefont {E.~M.}\ \bibnamefont {Stagar}}, \bibinfo {author} {\bibfnamefont {A.}~\bibnamefont {Simmons}}, \bibinfo {author} {\bibfnamefont {M.}~\bibnamefont {Nieberding}}, \bibinfo {author} {\bibfnamefont {A.}~\bibnamefont {Ayers}},\ and\ \bibinfo {author} {\bibfnamefont {C.}~\bibnamefont {Orban}},\ }\bibfield  {title} {\bibinfo {title} {{A controlled study of virtual reality in first-year magnetostatics}},\ }in\ \href@noop {} {\emph {\bibinfo {booktitle} {Physics Education Research Conference 2019}}},\ \bibinfo {series and number} {PER Conference}\ (\bibinfo {address} {Provo, UT},\ \bibinfo {year} {2019})\BibitemShut {NoStop}%
\bibitem [{\citenamefont {Madden}\ \emph {et~al.}(2018)\citenamefont {Madden}, \citenamefont {Won}, \citenamefont {Schuldt}, \citenamefont {Kim}, \citenamefont {Pandita}, \citenamefont {Sun}, \citenamefont {Stone},\ and\ \citenamefont {Holmes}}]{Madden2018VirtualBeyond}%
  \BibitemOpen
  \bibfield  {author} {\bibinfo {author} {\bibfnamefont {J.~H.}\ \bibnamefont {Madden}}, \bibinfo {author} {\bibfnamefont {A.~S.}\ \bibnamefont {Won}}, \bibinfo {author} {\bibfnamefont {J.~P.}\ \bibnamefont {Schuldt}}, \bibinfo {author} {\bibfnamefont {B.}~\bibnamefont {Kim}}, \bibinfo {author} {\bibfnamefont {S.}~\bibnamefont {Pandita}}, \bibinfo {author} {\bibfnamefont {Y.}~\bibnamefont {Sun}}, \bibinfo {author} {\bibfnamefont {T.~J.}\ \bibnamefont {Stone}},\ and\ \bibinfo {author} {\bibfnamefont {N.~G.}\ \bibnamefont {Holmes}},\ }\bibfield  {title} {\bibinfo {title} {{Virtual Reality as a Teaching Tool for Moon Phases and Beyond}},\ }in\ \href {https://doi.org/10.1119/perc.2018.pr.Madden} {\emph {\bibinfo {booktitle} {PERC Proceedings}}}\ (\bibinfo {address} {Washington, DC},\ \bibinfo {year} {2018})\BibitemShut {NoStop}%
\bibitem [{\citenamefont {Perkins}\ \emph {et~al.}(2006)\citenamefont {Perkins}, \citenamefont {Adams}, \citenamefont {Dubson}, \citenamefont {Finkelstein}, \citenamefont {Reid}, \citenamefont {Wieman},\ and\ \citenamefont {LeMaster}}]{Perkins2006PhET:Physics}%
  \BibitemOpen
  \bibfield  {author} {\bibinfo {author} {\bibfnamefont {K.}~\bibnamefont {Perkins}}, \bibinfo {author} {\bibfnamefont {W.}~\bibnamefont {Adams}}, \bibinfo {author} {\bibfnamefont {M.}~\bibnamefont {Dubson}}, \bibinfo {author} {\bibfnamefont {N.}~\bibnamefont {Finkelstein}}, \bibinfo {author} {\bibfnamefont {S.}~\bibnamefont {Reid}}, \bibinfo {author} {\bibfnamefont {C.}~\bibnamefont {Wieman}},\ and\ \bibinfo {author} {\bibfnamefont {R.}~\bibnamefont {LeMaster}},\ }\bibfield  {title} {\bibinfo {title} {{PhET: Interactive Simulations for Teaching and Learning Physics}},\ }\href {https://doi.org/10.1119/1.2150754} {\bibfield  {journal} {\bibinfo  {journal} {The Physics Teacher}\ }\textbf {\bibinfo {volume} {44}},\ \bibinfo {pages} {18} (\bibinfo {year} {2006})}\BibitemShut {NoStop}%
\end{thebibliography}%

\end{document}